\pdfoutput=1

\documentclass[letterpaper]{sig-alternate}

\usepackage[letterpaper=true,pdfview=FitH,pdfstartview=FitH,bookmarks=false]{hyperref}

\usepackage{graphicx}
\usepackage{balance}  % for  \balance command ON LAST PAGE  (only there!)
\usepackage{amsmath,amsfonts}
\usepackage{xspace}
\usepackage{ifthen}
\usepackage{url}
\usepackage{balance}  % for  \balance command ON LAST PAGE  (only there!)
\usepackage{color}
\usepackage{cite} % for sorting and compressing sequences of numeric citations
\usepackage{enumerate}

\usepackage{times} % Similar to default computer modern, but a bit smaller
\usepackage{transfercommands}
\newboolean{isDoubleBlind}
\setboolean{isDoubleBlind}{false}
\newcommand{\singleDoubleBlind}[2]{\ifthenelse{\boolean{isDoubleBlind}}{#2}{#1}}

\begin{document}
%
% --- Author Metadata here ---
\conferenceinfo{CIKM}{'12 Maui, Hawaii USA}
%\CopyrightYear{2007} % Allows default copyright year (20XX) to be over-ridden - IF NEED BE.
%\crdata{0-12345-67-8/90/01}  % Allows default copyright data (0-89791-88-6/97/05) to be over-ridden - IF NEED BE.
% --- End of Author Metadata ---

\title{Scaling Multiple-Source Entity Resolution using \\ Statistically Efficient Transfer Learning}

\numberofauthors{3} %  in this sample file, there are a *total*
\author{
% 1st. author
\singleDoubleBlind{%
\alignauthor
Sahand Negahban\titlenote{Research performed while SN was at Microsoft Research.}\\
       \affaddr{MIT}\\
       \email{sahandn@mit.edu}
% 2nd. author
\alignauthor
Benjamin I. P. Rubinstein\\
       \affaddr{Microsoft Research}\\
       \email{berubins@microsoft.com}
% 3rd. author
\alignauthor Jim Gemmell\\
       \affaddr{Microsoft Research}\\
       \email{jim.gemmell@gmail.com}%
}{\vspace{4em}}
}

\maketitle

\begin{abstract}
We consider a serious, previously-unexplored challenge facing almost all approaches to scaling up entity resolution (ER) to multiple data sources: the prohibitive cost of labeling training data for supervised learning of similarity scores for each pair of sources. While there exists a rich literature describing almost all aspects of pairwise ER, this new challenge is arising now due to the unprecedented ability to acquire and store data from online sources, features driven by ER such as enriched search verticals, and the uniqueness of noisy and missing data characteristics for each source. We show on real-world and synthetic data that for state-of-the-art techniques, the reality of heterogeneous sources means that the number of labeled training data must scale quadratically in the number of sources, just to maintain constant precision/recall. We address this challenge with a brand new transfer learning algorithm which requires far less training data (or equivalently, achieves superior accuracy with the same data) and is trained using fast convex optimization. The intuition behind our approach is to adaptively share structure learned about one scoring problem with all other scoring problems sharing a data source in common. We demonstrate that our theoretically motivated approach incurs no runtime cost while it can maintain constant precision/recall with the cost of labeling increasing only linearly with the number of sources.
\end{abstract}

%A category including the fourth, optional field follows...
%\category{H.3.3}{Information Systems}{Information Search and Retrieval}
%\category{D.2.8}{Software Engineering}{Metrics}[complexity measures, performance measures]

\section{Introduction}\label{sec:intro}

In this paper we investigate a serious and
previously-unexplored challenge to scaling joint entity
resolution (ER) to multiple sources: that of intractable labeling costs required to model
heterogeneities in real-world data sources.

\begin{figure*}[t!]
\begin{center}
\includegraphics[width=.7\linewidth]{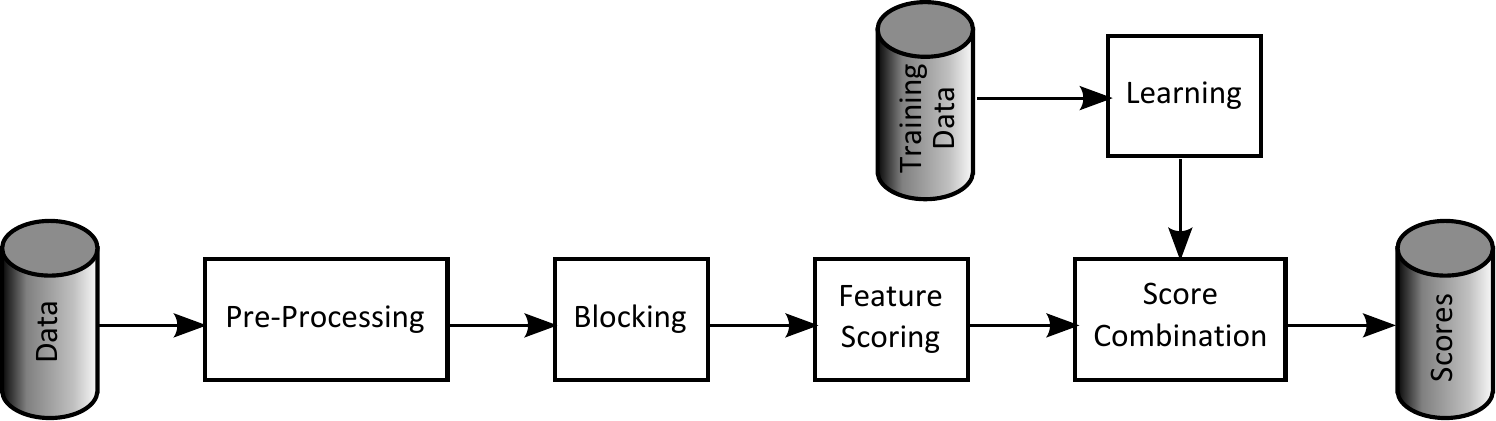}
\caption{A typical ER data flow. This paper focuses on the learning step, under multiple data sources, that infers the similarity of each pair of entities from attribute similarities, based on human-labeled examples of matching and non-matching record pairs.}
\label{fig:flow-chart}
\end{center}
\end{figure*}

Significant attention has already been focused on ER in the DB, data mining and statistics
communities, where the typically-stated goals are
computational performance (good runtime) and statistical performance (good
precision/recall)---\cf \eg \cite{Kopcke2010,DedupSurvey07} and references therein
for general discussions on ER. The most common approach for achieving good precision/recall
is to employ supervised learning to combine domain-expert-selected feature scores
into overall similarity scores~\cite{QDBMUD08-STEM,InfoSys-MultClassSys,Kopcke2010,ActiveLearn-KDD02,KDD03-MARLIN,ActiveAtlas-KDD02,JCoopInfoSys,KDD98-logistic,SIGMOD09-logistic}.
Indeed a recent, comprehensive evaluation of over 20 state-of-the-art ER systems~\cite{Kopcke2010a}, K\"opcke \etal found that on most tasks supervised learning-based matchers
offer superior performance.

However K\"opcke \etal also noted that statistical performance
comes at the price of human effort in labeling training examples, and explicitly highlight 
\emph{labeling cost as a key measure of matcher performance.} But while there have been
studies on multiple-source ER, and there are numerous applications in science, technology and 
medicine motivating effective approaches for ER
over multiple sources~\cite{StorWill05,Kelman08112003,Lawson06}, we are the first to note that \emph{state-of-the-art ER approaches have
intractable labeling cost on multiple sources}. Indeed to maintain constant precision/recall, we 
show that existing approaches suffer labeling costs that scale quadratically
as the number of sources increase.%
\footnote{We focus on the more general \emph{pairwise} matching problem
  as opposed to easier matching of multiple sources to a single master.}
The need for learning individual score functions when
faced with data heterogeneity has been explicitly~\cite{SourceAware07} and implicitly~\cite{QDBMUD08-STEM} acknowledged previously (\cf related work Section~\ref{sec:related} for more); however we are the first to comprehensively quantify 
this requirement. Finally, just as computational scaling can be tackled via cloud computing,
one may look to human computation (\eg via Amazon Mechanical Turk) to
address the labeling cost challenge. However, very many ER problems involve integrating highly
privacy-sensitive, or trade-secret, data that cannot be outsourced.

Our negative results on state-of-the-art approaches would suggest an impossible 
trade-off between precision/recall and labeling cost when performing ER on even
a moderate number of real-world, heterogeneous data sources. To address this problem,
we develop a brand new transfer learning algorithm that jointly learns to score pairs of data sources
while adaptively sharing common patterns of data quality. Training our algorithm \transfer involves
solving a convex optimization program via fast state-of-the-art composite gradient
methods~\cite{Nesterov07}. Motivated by a multiple-source ER problem
for the movies vertical in a major Internet search engine, we demonstrate both on a large real-world
movie entity crawl dataset (with sources 10x larger than any considered in \cite{Kopcke2010a})
and a large-scale synthetic dataset, that our \transfer algorithm
is superior compared to state-of-the-art approaches while incurring a \emph{labeling cost that
is only linear in the number of sources being resolved}. While this constitutes a major contribution
to entity resolution, \transfer is also of independent interest as a novel contribution to machine 
learning research as it leverages a previously-unseen pairwise structure between learning tasks
that is motivated directly by the application to ER.%
\footnote{Existing transfer learning approaches suit only the simpler multiple-source
  ER problem of all-against-master (as opposed to pairwise). \transfer formally subsumes
  such approaches.}

\lowKeyParagraph{Organization}
In Section~\ref{sec:problem} we present a precise problem statement and elaborate
on our running movie matching example. We then
develop the \transfer learning algorithm for low-labeling-cost multiple-source ER
in Section~\ref{sec:algo}.  
Sections~\ref{sec:exp} and~\ref{sec:results} present thorough
experimental evaluations on both real-world and synthetic
data. Finally we discuss related work in Section~\ref{sec:related} and conclude with directions
for future work in Section~\ref{sec:conc}. 

\lowKeyParagraph{Notation} On vectors $v \in \real^\pdim$, we let the
$\ell_q$ norm for $q \geq 1$ be defined as $\|v\|_q \defn
(\sum_{j=1}^\pdim |v_j|^q)^{1/q}$, and $\|v\|_\infty =
\max_i |v_i|$. We let $\sign(v)$ denote the
vector of the same dimensions whose $i^{th}$ element is the sign of
$v_i$ or $v_i/|v_i|$ if $v_i \neq 0$, and is equal to zero
otherwise.

\section{Problem Statement}\label{sec:problem}

%In this section we provide a detailed description of our problem
%and specify the model that we will assume throughout.

%\subsection{Scoring Function}

We now formalize our problem, which is to
produce functions that combine $\pdim$ similarity feature scores
$g(\entity_i, \entity_j)\in\real^\pdim$ between two
entities $\entity_i$ and $\entity_j$ taken from their respective sources
$\source_i$ and $\source_j$. As is common in ER, the feature scores
are typically chosen by a domain expert; the output of the combination
represents an overall similarity score between the entities that should achieve
strong precision and recall. We consider  $\numsources>2$ sources, and so
$i\neq j$ will be taken from
$\{1,\hdots,\numsources\}$ unless stated otherwise. As we shall demonstrate
empirically, automatically learning
the combination of feature scores typically requires prohibitively large amounts
of labeled training data for large $\numsources$.

\begin{df}
  The formal goal of the \term{Multi-Source Similarity Learning
    Problem} is: for each pair of sources $(i,j)$, learn a similarity
  scoring function $\func_{ij}$ mapping feature space attributes
  $g(\entity_i,\entity_j) \in \real^\pdim$ to a real-valued
  score. Negative (non-negative) scores are interpreted as predictions
  by $f_{ij}$ that a pair of entities is non-matching (matching), and
  the magnitude of the scores corresponds to a measure of confidence
  in the predictions. We desire to learn $\func_{ij}$ that achieve strong
  precision and recall using few labeled examples.
\end{df}

Figure~\ref{fig:flow-chart} depicts a typical ER system producing scores which can be fed
into subsequent merge or clustering functions~\cite{Kopcke2010}; resulting scores are
typically thresholded to produce the resolution. Prior to feature
scoring and score combination, the entities are normalized in a pre-processing step
(\eg in movie matching, producing clean movie titles, cast, directors, release years and runtimes);
and then blocking is employed to prune the pairs of entities considered in scoring, via
a linear pass hashing entities to blocks (\eg movie entities are hashed to their
non-stop-title-words so that only movies with a rare word in common are scored). In the movie matching example, feature scoring may produce title edit distance, year \& runtime absolute
difference, and Jaccard coefficients for cast and directors; then score combination computes overall
scores after learning how to do so from a human-labeled set of matching/non-matching entity pairs.

\section{Transfer Learning Algorithm}\label{sec:algo}

The primary goal of machine learning approaches is
statistical efficiency, formalized by the notion of a learning algorithm's
\term{sample complexity}: the amount of training data required for a desired
accuracy (with high confidence).
The \term{transfer learning paradigm}~\cite{Jalali10,Liu08,HuaZha09,zhang} has enjoyed recent interest in the machine learning
and statistics communities, due to its general principle of exploiting information gleaned
in multiple related learning tasks to reduce the tasks' sample complexities.
This section develops a new transfer learning algorithm for the Multi-Source Similarity
Learning Problem. As well as contributing a solution to the seemingly intractable 
labeling cost of performing ER over multiple sources, our algorithm \transfer represents
a contribution to machine learning research as it presents an approach to a novel
transfer learning problem with a unique inter-task structure.

\begin{figure}[t]
\begin{center}
\includegraphics[width=.7\linewidth]{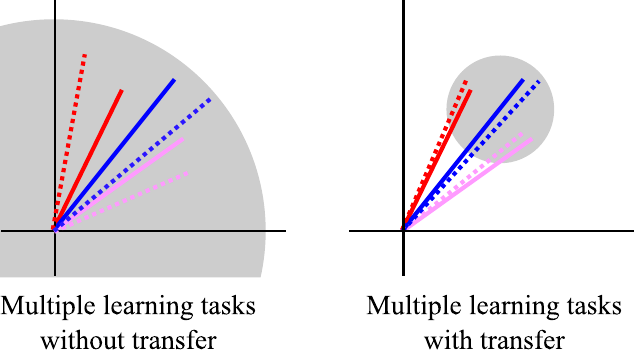}
\caption{Learning can be viewed as approximating a true concept (solid vectors) with a model (dotted vectors) taken from a class of models (grey ball). Left, three learning tasks (in red, blue \& pink) are typically performed independently. Task data is not pooled, and a large class of models may be consistent with the observed data. Right, transfer learning jointly learns a model class common to the tasks thereby implicitly sharing task data and effectively reducing the model class complexity.}
\label{fig:why-transfer}
\end{center}
\end{figure}

We now briefly overview the intuition behind transfer learning approaches in general.
In one setting of transfer learning we may
consider the problem of first performing one learning task (or several)
and then using the obtained
information to make a new learning task more efficient. In another
setting, we have multiple tasks that we wish to
learn from \emph{simultaneously} in the hopes that jointly learning
all models will result in a net decrease in sample complexity.
A common characteristic that each of these
settings share is that we wish to learn statistically
independent tasks. Each of these tasks represent separate problems
that share some common structure: either shared
support~\cite{Liu08,HuaZha09} or shared subspaces~\cite{zhang}.
Figure~\ref{fig:why-transfer} depicts the intuition of how transfer
learning improves accuracy via the latter approach. This figure
represents the setting in which the learning tasks' inferred models
(here vectors) should be highly clustered. This common structure
allows us to learn from the available information more effectively by
considering the problems jointly rather than separately.

In all forms of learning, the chosen classifier is taken from some
subspace of models that depends on parameters to the learning
algorithm, and the training data. With more data, the class of models
can be tightened to yield a more precise classifier. In low-data
settings, this is difficult to achieve. Transfer learning approaches
jointly learn the model class common to many learning tasks, while
learning each individual task's classifier. In so-doing, \emph{transfer
learning is able to learn accurately with less data.} A key element to
designing a successful transfer learning scheme is to appropriately
constrain the structure of the model class to reflect the shared
properties of the tasks' true classifiers. For example, the shared structure
in Figure~\ref{fig:why-transfer} is depicted as the true
classifiers sitting in a small Euclidean ball.

%\todo{move? contribution: scaling via transfer}
%One of the contributions of this paper is the application of the
%jointly-transfer learning paradigm to improve the scalability of
%labeling requirements in entity resolution. Namely, as we discussed in
%the previous section, a problem of interest is in learning a method to
%score a pair of entities from two distinct sources. In general, we may
%have a vast number of sources, and we would like to simultaneously
%learn a scoring function for all pairs of sources.

A challenge that arises in our setting of tasks corresponding to learning source-pair similarity 
functions $\{\func_{ij}\}$ is in handling the interactions
between the sources $\{S_i\}$. For example, a standard transfer learning approach to learning a
scoring function between sources $A$ and $B$ and between $A$ and $C$
would be to treat these two tasks just as it would third task for $D,E$, ignoring
the fact that some tasks share a common source: here $A$. A new model
would allow us to more accurately learn a scoring function across pairs of sources
for which no available training examples exist.

%\todo{delete? mechanics}
%One question that arises is rather than learning a separate function
%for each pair $i,j$, we could instead learn a single function $f$ that
%maps pairs of entities \emph{and} pairs of sources to a score. This
%method is another viable technique to jointly learn our scoring
%functions, and indeed represents a much broader class of possible
%scoring functions. Namely, given entities $i_1,i_2$ and a pair of
%sources $S_1,S_2$, then we can set $f(i_1,i_2,S_1,S_2) =
%f_{S_1,S_2}(i_1,i_2)$, which shows that the class of functions that we
%wish to consider is equivalent to the above class. However, as we
%develop our method and model in Section~\ref{sec:models}, we will
%reduce the complexity of the function class in order to make the
%learning problem both statistically and computationally efficient.

\subsection{Learning Models}

We begin to derive our new transfer learning algorithm for ER by expressing
the class of models our algorithm will learn over. For reasons made clear below,
we design \transfer to learn linear classifiers; however we later compare this
approach against state-of-the-art non-linear algorithms, and we note that the
techniques described here are general and can be kernelized to produce non-linear
analogues.

\subsubsection{Linear Classifiers in ER}

Specifying an appropriate
model allows us to avoid overfitting the training data. However, 
as the complexity of our model increases so too does the number of 
training examples---the sample complexity---required to fit all of
the available ``degrees of freedom'' of our model.
Hence, we will need to take the amount of available training data
and the learning task at hand into consideration when specifying our model. 

%We recall that
%while our general problem of interest is finding a scoring function,
%our motivating example is that of entity resolution.
In ER~\cite{JCoopInfoSys,KDD98-logistic,InfoSys-MultClassSys,QDBMUD08-STEM,SIGMOD09-logistic},
and many other problems~\cite{Lawson06,SreRenJaa04}, it has been shown that linear models perform exceptionally well for explaining the
behavior between feature score vectors $x$ and output labels $\yval$.
The choice of a linear model serves a dual statistical and computational
purpose. Linear models can be evaluated very quickly and are
also inexpensive to store, requiring only $\pdim+1$ doubles---together
making the model ideal for large-scale learning.  From a
statistical perspective, given enough features, we can accurately
model the interactions in our data.  
% Recent
%advances in high-dimensional statistics have shown that real-world data 
%can often be fit by considering models with enormous numbers of features,
%where only a small subset are chosen as relevant.
%Linear models arise in
%medical research~\cite{Lawson06}, entity %,Erlich09
%resolution~\cite{JCoopInfoSys,KDD98-logistic,InfoSys-MultClassSys,QDBMUD08-STEM,SIGMOD09-logistic}, and collaborative
%filtering~\cite{SreRenJaa04}.
Formally, we assume that a given input set of features $x$ and an output label $\yval$ can be related by
\begin{equation*}
  \yval = \sign(\inprod{\wvec}{x} + \noisevec)\;,
\end{equation*}
where $\noisevec$ is a bias term capturing the fact that the model is
not exact due to noise. Here $\wvec$ acts as weight vector, placing
varying importance on each of the feature scores in $x$. The setting
of $\wvec$ results in splitting our feature score space into two
half-spaces, since we have two classes. Finally, we will take our
similarity scoring function for a given source pair $(i,j)$ to be\footnote{We
drop the term $\noisevec$ for convenience, without loss of generality.}
\begin{equation*}
  \func_{ij}(x) = \inprod{\wvec_{ij}}{x}.
\end{equation*}
Hence the Multi-Source Similarity Learning Problem corresponds
to inferring the weight vectors $\wvec_{ij}$.

We will compare the transfer learning approach based on linear models
of this section, with both linear and non-linear state-of-the-art baselines
in Sections~\ref{sec:exp} and~\ref{sec:results}.
There is a trade-off: on the one hand, more features allow us to model more
interactions, on the other hand, more features result in a more
complex model that can be susceptible to overfitting, at a detriment to
sample complexity \ie labeling cost.

\subsubsection{Transfer Learning Model}

%Obtaining training data can be prohibitively expensive, as noted by researchers
%studying active learning approaches to reducing sample complexity when learning
%a single similarity function~\cite{ActiveLearn-KDD02} and a recent comprehensive study 
%highlighting labeling as a key cost of ER~\cite{Kopcke2010a}.
%In the setting of
%ER and related problems, we typically require humans to consider
%two sources and
%explicitly locate the matching pairs between the sources, which can require
%significant investments in time and domain expertise. Therefore, reducing the
%required amount of training examples---the sample complexity---while retaining
%good statistical performance is greatly important.

While we have a number of separate tasks across different
pairs of sources---naively leading to a quadratic scaling of the sample complexity
with the number of sources $\numsources$---one training example from source
pair $(i,j)$ could 
inform learning to score another source pair $(j,h)$ and conceivably even $(h,k)$.
This intuition motivates our interest in applying transfer learning to uncover 
shared characteristics between the different source pair tasks. As borne out in 
our experiments, doing so will
effectively allow us to share examples across many different source
pairs in order to most efficiently use the available resources and successfully scale
to resolving multiple heterogeneous data sources.

Two extreme forms of transfer are in common use in practice today: either learn
each classifier $\func_{ij}$ separately (so as to model heterogeneities in the sources 
at a great labeling cost), or pooling all available data and learning a single classifier
(mitigating the labeling cost at the expense of flexibility). Both existing approaches
represent two extreme forms of transfer (none and complete transfer, respectively).
An ideal method should behave
between both extremes and allow the data to dictate the most appropriate
behavior. When there is very limited data, we may not have
enough information to describe the difference in characteristics between sources. 
As we gain more information, our method should adapt
and take into account any added information. To that end, we introduce
a method that we call \emph{transfer}.  For this
model, we assume that our weight vectors decompose as
\begin{eqnarray}
  \label{EqModel}
  \wvec_{ij} & = & \wvec_0 + \wvec_i + \pairvec_{ij}\;,
\end{eqnarray}
where the vector $\wvec_0$ captures the general trends, for example,
movies with the same casts are generally going to be similar. The
weight vector $\wvec_i$ accounts for the specific effects induced by
the particular source and the vector $\pairvec_{ij}$ handles the
pairwise deviations and can also be applied to guarantee that
$\wvec_{ij} = \wvec_{ji}$.

\subsection{Regularized Learning Formulation}\label{sec:learning}

We now formulate an optimization program for learning the underlying
pairwise score functions.
There has been a flurry of research in
developing efficient techniques for finding parameters that can
accurately describe the data using models as those described above. A
number of techniques are based on optimizing a convex function for
efficiently recovering the parameters.  Such convex programs have seen
tremendous theoretical and experimental success in the
literature~\cite{CandesTao05,Geer08,Liu08}.

Before proceeding, we
recall that the sources are indexed by an integer in $\{1,\ldots,\numsources\}$
so that (abusing notation) $\source \in
\{1,\hdots,\numsources\}$. Furthermore, 
we will let $(i(k),j(k))$ denote the source pair that the
$k^{th}$ example was drawn from. Given that, we write our $k^{th}$
training example as
$(\featvec_k,\source_{1,k},\source_{2,k},\yval_k)$, where $\featvec_k
= g(\entity_{1,k},\entity_{2,k})$ denotes the feature vector
representation of the pair of entities
$(\entity_{1,k},\entity_{2,k})$, $\source_{1,k}$ and $\source_{2,k}$ 
represent the source indices of the entities, and $\yval_k$ represents
the true label. With this notation, we may
propose to learn an Equation~\eqref{EqModel} model that solves the
convex program
\begin{multline}
  \label{EqMest}
  \argmin_{\wvec_0,\wvec_i,\pairvec_{ij}} \underbrace{\frac{1}{2} \sum_{k=1}^\numobs (\yval_k - \inprod{\wvec_0+\wvec_{\sourcea(k)} + \pairvec_{\sourcea(k)\sourceb(k)}}{\featvec_k})^2}_{\text{empirical risk term}}\\
 \ \ + \, \underbrace{\regpara \sum_{i=1}^\numsources \| \wvec_i\|_1 +
    \frac{\regparb}{2} \sum_{i,j}
    \|\pairvec_{ij}\|_2^2}_{\text{regularization terms}}\\
  \text{s.t. $\wvec_0 + \wvec_i + \pairvec_{ij} = \wvec_0 + \wvec_j +
    \pairvec_{ji}$}
\end{multline}
The result of this program are estimates of our weight vectors $w_{0}$, $w_i$, and
$\pairvec_{ij}$. We now take a moment to discuss 
Program~\eqref{EqMest}.
The objective function can be decoupled into
two components: an empirical risk or loss term and a regularization
term.

\lowKeyParagraph{Loss Term}
The loss term aims to encourage predictions on the training input
feature vectors to match the training labels.  Furthermore, we note
that our assumption on the form of the pairwise score functions is
built into the optimization procedure. That is, $\inprod{\wvec_0 +
  \wvec_{i} + \pairvec_{ij}}{\featvec_k}$ should be close to
$\yval_k$. %Such a model has been shown to behave well for
%classification problems, and we will provide experiments in the sequel
%that confirms this.
We note that there are other alternative options
for the loss term such as those used in the logistic regression or support vector
machine linear models, both of which are also used extensively in the
literature. 

\lowKeyParagraph{Regularization Terms}
While the empirical loss term encourages our parameters to closely fit
the model, the regularization terms exist in order to penalize
overly-complex models and avoid overfitting. For the regularization we
penalize the source weight vectors $\wvec_i$ by the $\ell_1$ norm and
the pairwise weight vectors $\pairvec_{ij}$ by the $\ell_2$ norm
squared. These choices have both been extensively studied in the
literature due to a number of desirable consequences that they each
have.
% \todo{cite}.
It has been shown that the $\ell_1$ norm encourages
solutions to convex optimization procedures to be sparse (the $\ell_1$
norm essentially acts as a convex surrogate to the $\ell_0$ norm, or
the total number of non-zero parameters in a vector). Authors have
established both theoretical results and experimental results
demonstrating the performance of the $\ell_1$
norm~\cite{Tibshirani96,CandesTao05, Donoho06}.  By encouraging the
$\wvec_{i}$ to be sparse, we capture the fact that each source (for the most
part) should behave as the nominal source represented by $\wvec_0$. This
type of assumption has also appeared in the context of robust regression~\cite{Jalali10}
and low-rank sparse matrix decompositions~\cite{Chand11}.
The $\ell_2$ norm squared terms acts to restrict the size of $\pairvec_{ij}$ (without
necessarily requiring sparsity), allowing pairwise perturbations away from the
nominal behavior between two sources but avoiding overfitting~\cite{KernelBook}.
This choice also accounts for the fact that in general $\wvec_0 + \wvec_i$ will not
in general equal $\wvec_0 + \wvec_j$.

\lowKeyParagraph{Parameter Selection}
We may tune
the parameters $\regpara$ and $\regparb$ to achieve various levels of
model complexity and control the amount of transfer. These parameters can be selected via extending existing
theoretical results in the literature~\cite{BiRiTsy08} or based on a
user's prior knowledge for the problem. Another popular method (adopted
in this paper) is to
apply cross validation, and use a hold-out set of the data to 
select the parameters~\cite{lars}.

\lowKeyParagraph{Extending the Learning Algorithm}
Our construction allows for a number of choices
for the empirical risk and regularization functionals, and we found that our current choices worked
well practically from a statistical perspective as well as a
computational one. It would be a relatively trivial task to 
modify our optimization to be more like a pairwise transfer learning version
of other linear model-based learners such as logistic regression or
SVMs. \emph{Our contribution is a generic transfer learning approach for
ER which encompasses a family of algorithms; one of which we focus on here
as a first study on using transfer in multiple-source ER.}

\subsection{The Algorithm}

We now proceed to derive methods for solving  Program~\eqref{EqMest}.

\subsubsection{Optimality Conditions}

The following result derives from applying the Karush-Kuhn-Tucker (KKT)
conditions that govern the program's optimality conditions~\cite{Boyd02}.

\begin{lem}
Suppose
that we are given optimal solutions to the convex Program~\eqref{EqMest}:
$\wstar_0$, $\wstar_i$, and $\pairstar_{ij}$, and that we let
\begin{eqnarray*}
  X^{ij} & \defn &\sum_{\{k | i(k)=i,j(k)=j\}} \featvec_k \featvec_k^T\;, \quad \text{and} \\
  g^{ij} & \defn &\sum_{\{k | i(k)=i,j(k)=j\}} \featvec_k \yval_k\;.
\end{eqnarray*}
Then an application
of the KKT conditions yields that
\begin{multline*}
  \pairstar_{ij} \; = (X + 2 \mu I)^{-1}(g - X \wstar_0 + \mu \wstar_j
  - (X + \mu I) \wstar_i)\;.
\end{multline*}
\end{lem}
We would like $\pairstar_{ij}$ to be as small as possible, which would
equate to setting $\mu$ as large as possible. Therefore, letting $\mu$
go to $\infty$, we have that
\begin{equation*}
  \pairstar_{ij} \; = \frac{1}{2} \left ( \wstar_j - \wstar_i \right ).
\end{equation*}
Therefore, we observe that under $\mu \to \infty$, each
$\pairstar_{ij}$ is simply half the difference between the source vectors
$\wstar_i$ and $\wstar_j$. Thus, we immediately obtain
\begin{equation*}
  \wstar_0 + \wstar_i + \pairstar_{ij} \; = \; \wstar_0 + \frac{1}{2} \left ( \wstar_i + \wstar_j \right ).
\end{equation*}
The above setting of $\mu$ and subsequent choice of $\pairstar_{ij}$
is used throughout the remainder of the paper. This
model is one that lends itself to far
simpler computation since we are only required to compute $\wvec_i$
for $0 \leq i \leq \numsources$. Furthermore, the model still captures
interesting characteristics of each of the sources while 
learning the common characteristics shared across all sources.
%%\section{Algorithm}

We now present our method for solving
Program~\eqref{EqMest}. Standard convex solvers can be
employed in the setting when the number of sources and the dimensionality
of the problem are small. However, as the number of sources $\numsources$
and the number of dimensions $\pdim$ both grow, we must rely on specialized
methods that can overcome the potential computational challenges
for solving such minimization programs. Before proceeding we define
the loss function to be
\begin{equation*}
  \loss(\wstar_0,\wstar_{i}) \; = \; \frac{1}{2} \sum_{k=1}^\numobs (\yval_k - \inprod{\wvec_0+\frac{1}{2} (\wvec_{\sourcea(k)} + \wvec_{\sourceb(k)})}{\featvec_k})^2
\end{equation*}
and the regularization terms to be
\begin{equation*}
  \reg(\wvec_{i}) \; = \; \regpara \sum_{i=1}^\numsources \|\wvec_i\|_1
\end{equation*}

% \begin{figure}[t]
% \begin{center}
% \includegraphics[width=1\columnwidth]{figures/soft-threshold.pdf}
% \caption{Soft threshold operator with parameter $\tau$.}
% \label{fig:soft-threshold}
% \end{center}
% \end{figure}

\subsubsection{Solution via Composite Gradient Methods}

We now develop a simple algorithm for solving our above convex program
based on \term{composite gradient descent} methods for optimizing composite objective
minimization problems~\cite{Nesterov07}. The method is an iterative
algorithm that updates the estimates at each time step based on the
current gradients. If we take $\wvec_i(s)$ to be the iterate at the $s^{th}$
iteration of the algorithm, then we have that
% \begin{algorithm}{Transfer}
%   $\wvec_0(s+1)$ \;
%   $\defn$ \; $\wvec_0(s) - \stepsize \nabla_{\wvec_0}\loss(\wvec_0(s),\wvec_i(s))$ \\
%   $\wvec_i(s+1)$ \;
%   $\defn$ \; $\softthresh{\stepsize \regpara}{\wvec_i(s) - \stepsize \nabla_{\wvec_i}\loss(\wvec_0(s),\wvec_i(s))}$,
% \end{algorithm}
\begin{eqnarray*}
  \wvec_0(s+1) &
  \defn& \wvec_0(s) - \stepsize \nabla_{\wvec_0}\loss(\wvec_0(s),\wvec_i(s)) \\
  \wvec_i(s+1) &
  \defn& \softthresh{\stepsize \regpara}{\wvec_i(s) - \stepsize \nabla_{\wvec_i}\loss(\wvec_0(s),\wvec_i(s))},
\end{eqnarray*}
where $\softthresh{\tau}{v}$ is the soft-thresholding operator defined on vector $v$,
with parameter $\tau$, as
\begin{equation*}
  \softthresh{\tau}{v} = \sign(v) \max(|v|-\tau,0) \enspace.
\end{equation*}
Here $\max(|v|-\tau,0)$ is the vector of the same
dimension as $v$ whose $i^{th}$ entry is the maximum of $|v_i| -
\tau$ and zero.
%(\cf Figure~\ref{fig:soft-threshold}).

Intuitively, we
update the vectors in the direction that will decrease the loss the
most.  In the case of $\wvec_i$, we must also account for the $\ell_1$
regularization terms, which result in truncation operations after
taking the gradient step.  The addition of an $\ell_1$ regularization
term makes the optimization procedure non-smooth.  While we could
employ second-order methods~\cite{Boyd02} for solving this problem, those would be
intractable for problems in higher dimensions. Hence, we rely on
first-order gradient-based methods for computational tractability.
The down-side of applying a gradient-based method is that optimizing
general non-smooth functions can become prohibitively slow. However,
it has recently been shown that optimizing based on composite
objective methods will result in iterates that converge
geometrically fast~\cite{AgaNegWai10}.

\section{Experiments}\label{sec:exp}

We next discuss experiments for
verifying the behavior of our transfer learning algorithm and to compare it against the
state-of-the-art. The results presented in Section~\ref{sec:results}
demonstrate significant gains on real-world movie matching and synthetic datasets,
showing that \transfer can achieve strong performance with low labeling cost that
scales only linear with the number of sources.

%In what follows we will present a detailed explanation of the
%evaluation methodology that we used for the experiments performed on
%our movie data and synthetic data. We will then move on to present the
%experimental results and offer insights into the observed
%behavior. Our first discussion will be based on understanding how
%varying the number of available training examples can alter the
%performance of the algorithm. We will then proceed to demonstrate on
%synthetic data how varying the number of sources alters the
%statistical behavior of the algorithm---source complexity. Finally, we will
%present timing results that exhibit favorable computational properties
%of our method.

\subsection{Baseline Approaches}\label{sec:models}

We consider three approaches representing the spectrum of state-of-the-art in
ER: pairwise and pooled linear classifiers (which as we argue are
actually special cases of transfer learning), and support vector machines (a 
non-linear learner popular in ER).

\lowKeyParagraph{Single}
The first model, called the \pooled or \term{\pooledalt} method, simply
assumes that $\wvec_{ij} = \wvec_0$ for all pairs of sources $i,j$ \ie
by constraining all the $\wvec_i = \pairvec_{ij} = 0$ in \transfer.
We have \emph{pooled} all of the tasks into a \emph{single} base task---we
essentially impose maximum transfer between each task. In this setting,
we are effectively required to only estimate $\pdim+1$ parameters, which
can be done very effectively with order $\pdim$ training
examples~\cite{NegRavWaiYu09}. Hence, we have greatly reduced the
model complexity of the problem at the expensive of ignoring any of
the unique behavior of individual sources.

\lowKeyParagraph{Pairwise Independent}
At the other extreme, the method \indep considers the
situation where all normal vectors $\wvec_{ij}$ are learned
without any shared characteristics.
This setting has no transfer as we make no assumption as
to the structure between the tasks of learning pairwise scoring
functions. The model complexity is prohibitively large since the number of
parameters to estimate scales as $\numsources^2 \pdim$. Hence, we
would require order $\numsources^2 \pdim$ training examples just to
learn each of the classifiers. However under heterogeneous sources, with
enough data, this approach should achieve far superior accuracy over \pooled.

\lowKeyParagraph{Non-linear}
Our third baseline model, denoted \svm, 
involves learning a non-linear function. Unlike the above linear models
that take a weighted sum of the pairwise attributes, the non-linear
learner will return an arbitrary function on the features.  So that \svm may
vary its output depending on the originating sources, it is natural to encode
the source pair identities in addition to the feature scores---the feature vector
presented is $(g(\entity_i),g(\entity_j),i,j)$ with source pair $(i,j)$ 
encoded into a length $\numsources$ vector that is all zeros except a
$1$ in the $i^{th}$ and $j^{th}$ components (since source ordering is
irrelevant). In this way \svm has the flexibility of modeling sources individually,
while implicitly transferring patterns learned between sources.
For our experiments we take \svm to be the support vector machine
(SVM) with Gaussian kernel, which corresponds to the most widely used and
flexible feature mapping. This SVM takes in two parameters: the cost
parameter $C>0$ and the kernel variance $\sigma>0$.

\begin{rem}
\pooled and \indep are both special (extreme) cases of \transfer
and are also of independent interest since together they represent one kind of
state-of-the-art technique---linear classification---that has enjoyed success
in ER~\cite{JCoopInfoSys,KDD98-logistic,InfoSys-MultClassSys,QDBMUD08-STEM,SIGMOD09-logistic}.
With adaptively selected parameters (via cross validation), we expect
\transfer to find a balance between the label-economical but inflexible
\pooled and the flexible but label-hungry \indep.

SVMs are regarded as one of the
most effective learners in ER~\cite{Kopcke2010a,Kopcke2010,KDD03-MARLIN,ActiveLearn-KDD02,FEBRL,Bilenko03,Hetero05,QDBMUD08-STEM}.
Like \transfer, the SVM can adapt its model complexity for the problem (via
its parameters). However to take advantage of its flexibility over
linear methods more data is typically needed. Moreover while it may learn different
classifications for different sources (as these are encoded in the feature vector),
and indeed \emph{transfer} between these tasks, the SVM does not possess the pairwise
structural knowledge that is built in to \transfer.
\end{rem}

\subsection{Algorithm Implementation}

For our experiments, we implement \transfer as described in Section~\ref{sec:algo} in the statistical
computing environment R. We implement the baseline \pooled and \indep
algorithms based off of the more general \transfer implementation; however to speed
up the baseline algorithms in-line with fast state-of-the-art
implementations---for fair and representative comparisons in our timing experiments---we
exploit standard computational tricks not available for the general transfer
learning case.

We use the R e1071 package's SVM routines, which are a wrapper for the popular
libSVM library, for implementing \svm. We employ 10-fold cross-validation
for selecting optimal SVM parameters $(C,\sigma)$ over a grid of candidates as is standard.

\subsection{Evaluation}

\begin{table}[t]
\begin{center}
\begin{tabular}{|cccccc|}
\hline
IMDB & AMG & Flixster & MSN & Netflix & iTunes \\
\hline 
526k & 306k & 141k & 104k & 76k & 12.5k \\
\hline
\end{tabular}
\caption{The number of entities per crawled movie data source.}
\label{tab:movie-data}
\end{center}
\end{table}

In order to investigate the statistical performance of the methods'
scores, we adopt a common threshold algorithm:  we declare
that two entities $\entity_1$ and $\entity_2$ are a match if their
score is above threshold $\thresh$, which we vary to
produce a set of potential classifiers. Therefore, given a scoring
function $\func$ and a set of examples $\{(\featvec_k,\yval_k)\}$, we aim
to compare the true labels $\yval_k$ against the estimated labels
\begin{equation*}
  \yhat_k = \sign(\func(\featvec_k)-\tau)\enspace.
\end{equation*}
We evaluate the performance of classifier's \emph{classifications} through precision and recall, defined
in the usual way as follows.
\begin{df}
  Let
  \begin{align*}
    TP & \defn \{k | \yhat_k = \yval_k = 1 \} \quad \text{be the set of true positives,}\\
    FP & \defn \{k | \yhat_k=1 \; \text{and} \; \yval_k=0 \} \quad \text{be the set of false positives,}\\
    FN & \defn \{k | \yhat_k = 0 \; \text{and} \; \yval_k =1 \} \quad \text{be the set of false negatives.}
  \end{align*}
  We define precision \precision and recall \recall in terms of these sets:
  \begin{align*}
    \precision  \defn \frac{|TP|}{|TP| + |FP|} & & \recall  \defn \frac{|TP|}{|TP| + |FN|}.
  \end{align*}
\end{df}
Hence, for varying threshold $\thresh$, we have a range of different
\precision and \recall values that together to form a precision-recall curve. We also
measure test error as follows, which combines both false positives and negatives.
\begin{equation}
  \label{EqProbErr}
  \proberr \defn \frac{1}{m} \sum_{k} \ind(\yval_k \neq \yhat_k).
\end{equation}
%Now suppose that we have $m$ examples and that example $k$ consists of
%a feature vector $\featvec_k$, label $\yval_k$, and a score
%$\scoreval_k$. Furthermore, suppose that we have estimated a score
%function $\func$, where we have suppressed the dependency on the pair
%of sources $i$ and $j$ for ease of notation. Then we let the mean
%squared error of the square values be
%\begin{equation*}
%  \mse(f) \defn \frac{1}{m} \sum_{k=1}^m (\func(\featvec_k)-\scoreval_k)^2.
%\end{equation*}

\subsection{Datasets and Pre-Processing}

We employed two large-scale datasets in our experiments.

\lowKeyParagraph{Real-World Move Data} Six major online movie
sources were crawled for use in
\singleDoubleBlind{the Bing movies vertical}{the movies vertical of a major Internet search engine}.
The number of records obtained are given in Table~\ref{tab:movie-data}. For each movie we
obtained its title and alternate titles, release year,
runtime, cast, and directors.  From these attributes we performed
basic string cleanup and blocked on common (non-stop) words in the titles.  Each
raw feature produced one feature score: Jaccard for titles, directors
and cast; and absolute difference for runtime and year.  Humans
labeled 200 entity pairs across each source pair. In our following
experimental results on this movie data, we learn the
scoring functions on various sources (as specified) but
evaluate precision and recall against movies from the pair IMDB
and iTunes. This choice was made in order to demonstrate the behavior
across a specific pair rather than averaging across all available
pairs. We held out a subset of the movie data as the
test set. We then used the remainder for training the methods.
In order to improve the conditioning of the problem, as is standard in machine learning,
we \term{standardized} the data by subtracting feature score means and 
dividing by standard deviation, making the features zero mean and unit variance
and so placing the features on equal footing.
%any feature was one. Next, we computed the mean of each feature in the
%training set and subtracted that amount from the data. Note that these
%values are all based on the training set, so we must save the values
%and apply similar operations to the test set when we are ready to
%evaluate the procedures. Note that we do not compute the scale and mean
%parameters on the entire data set, because the test-set would
%introduce a bias to our procedure.

\lowKeyParagraph{Synthetic Data} We synthesized raw true attributes
for each underlying latent entity uniformly at random in a unit
interval. Then each record representing an entity in a source was
produced by perturbing each of the attributes randomly with
low-variance Gaussian noise. Feature-level scores were then produced
using a simple difference between the attribute values of pairs of
entities. It is important to note that perturbing the feature-level
scores would be an incorrect methodology since the scores would not
observe any kind of triangle-inequality-like property as is the case
for ``real'' ER problems. We produced up to 30
synthetic sources to stress test the approaches, and used 10k test pairs
total.

\begin{figure}[t!]
\begin{center}
%\begin{minipage}[t]{0.48\textwidth}
\centering
\includegraphics[width=1\linewidth]{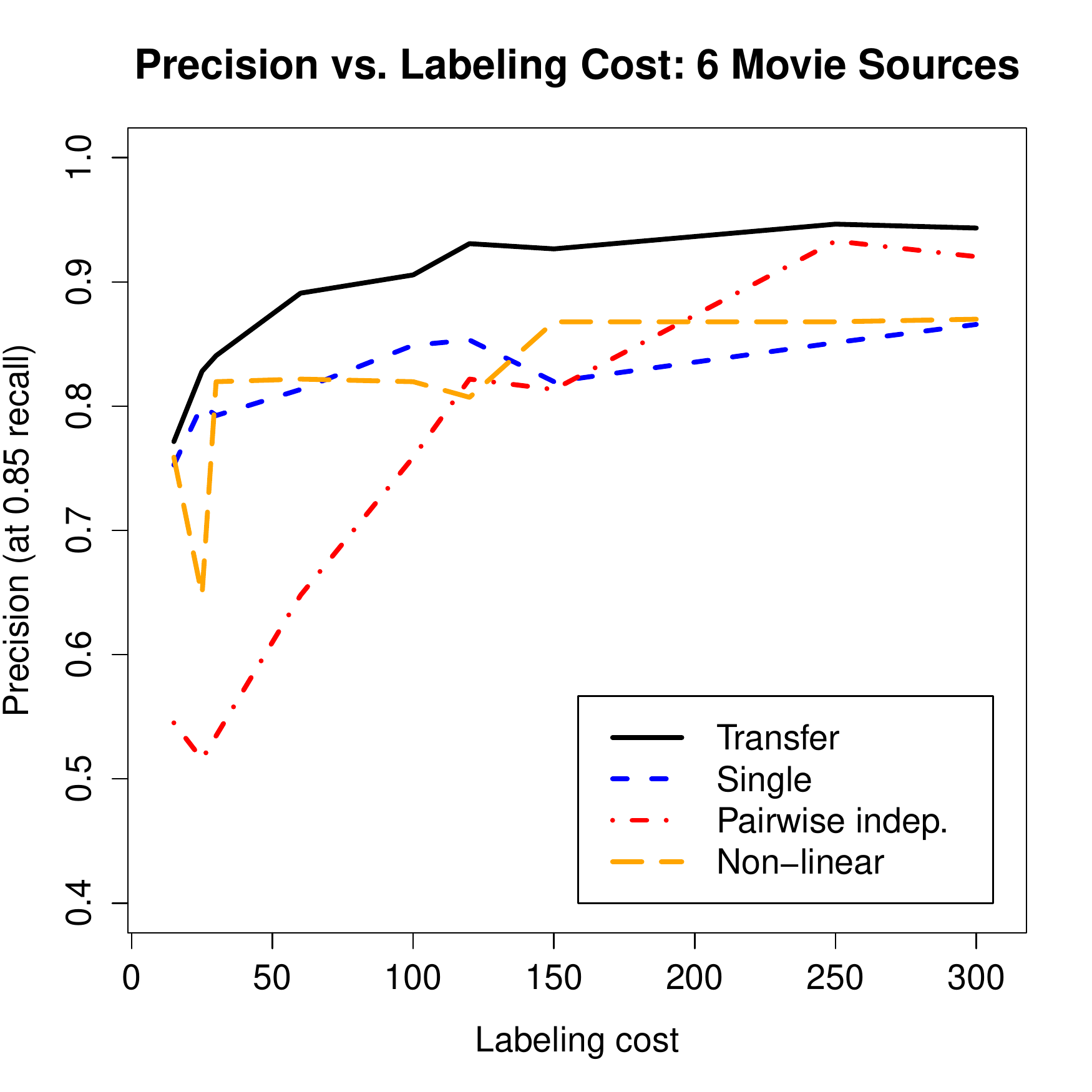}
%\end{minipage}\hfill
%\begin{minipage}[t]{0.48\textwidth}
%\centering
%\includegraphics[width=1\linewidth]{figures/sampleComplexity-trials50-mse.pdf}
%\end{minipage}
\caption{Precision achieved by each method on the 6 movie sources, at 0.85 recall, for varying number of examples.} %Plotting on the left axis (left) test error (right) mean squared error.}
\label{fig:prec-vs-labeling}
\end{center}
\end{figure}

\begin{figure*}
\begin{tabular}{l c r}
\includegraphics[width=.33\linewidth]{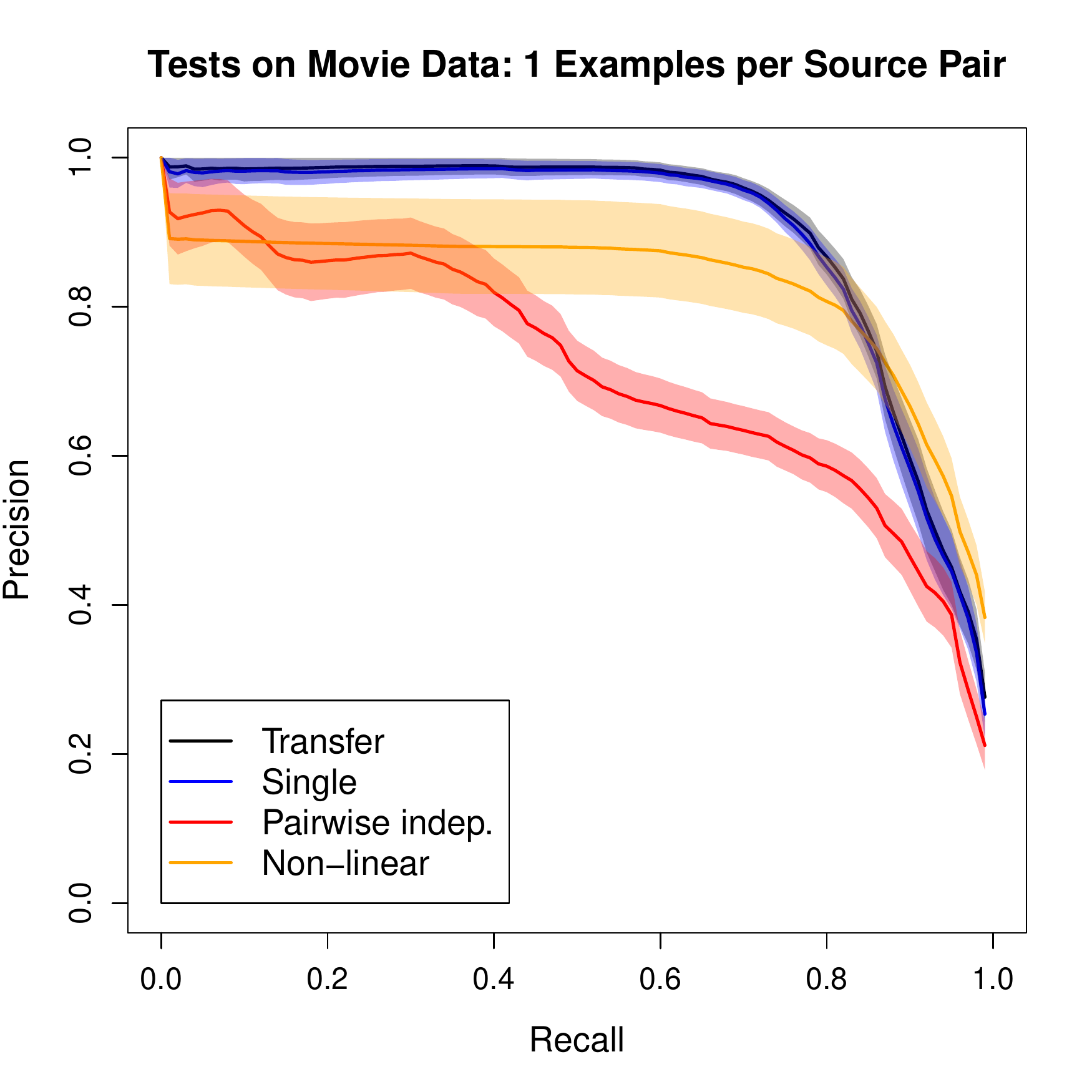} &
\includegraphics[width=.33\linewidth]{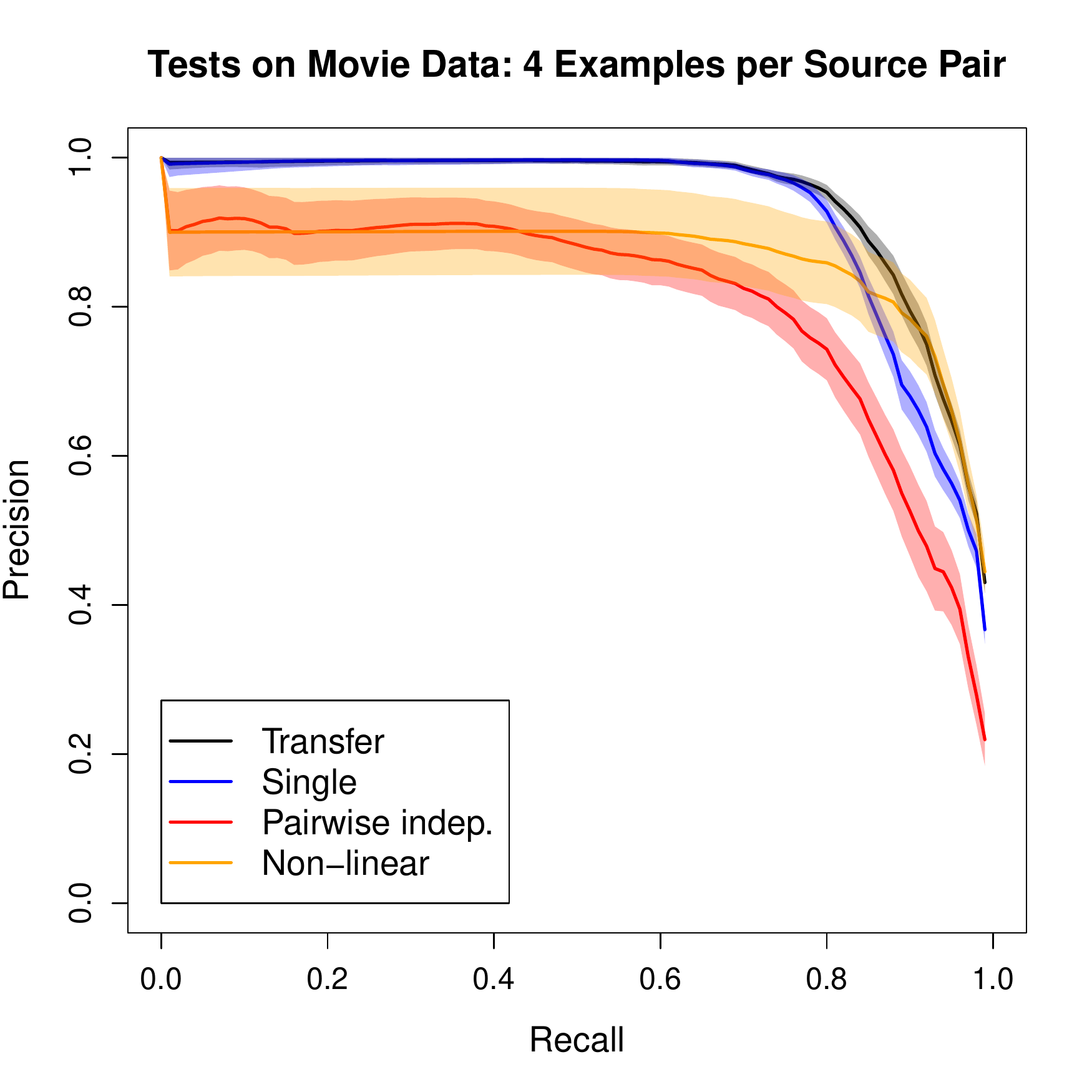} &
\includegraphics[width=.33\linewidth]{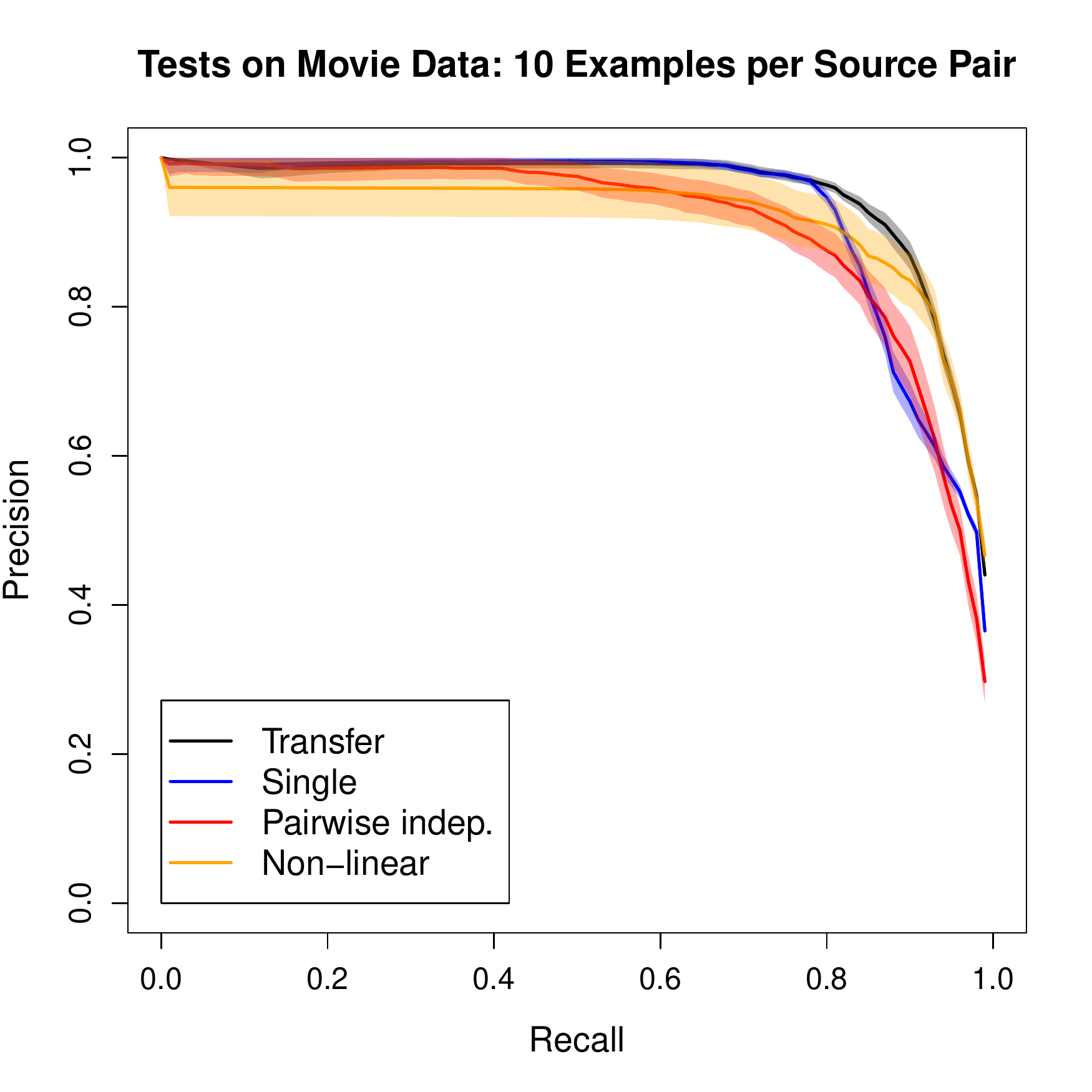}\\
\hline
\includegraphics[width=.33\linewidth]{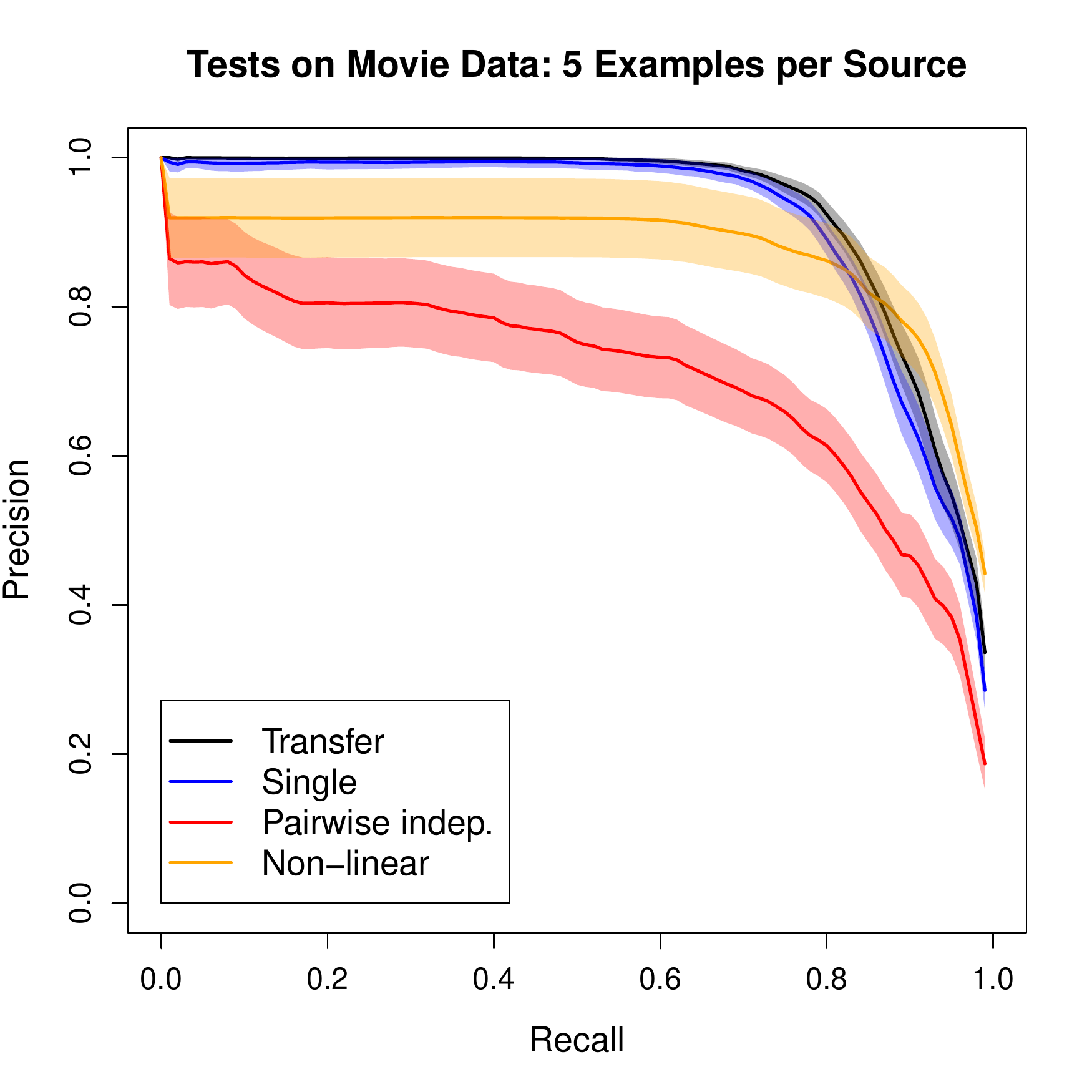} &
\includegraphics[width=.33\linewidth]{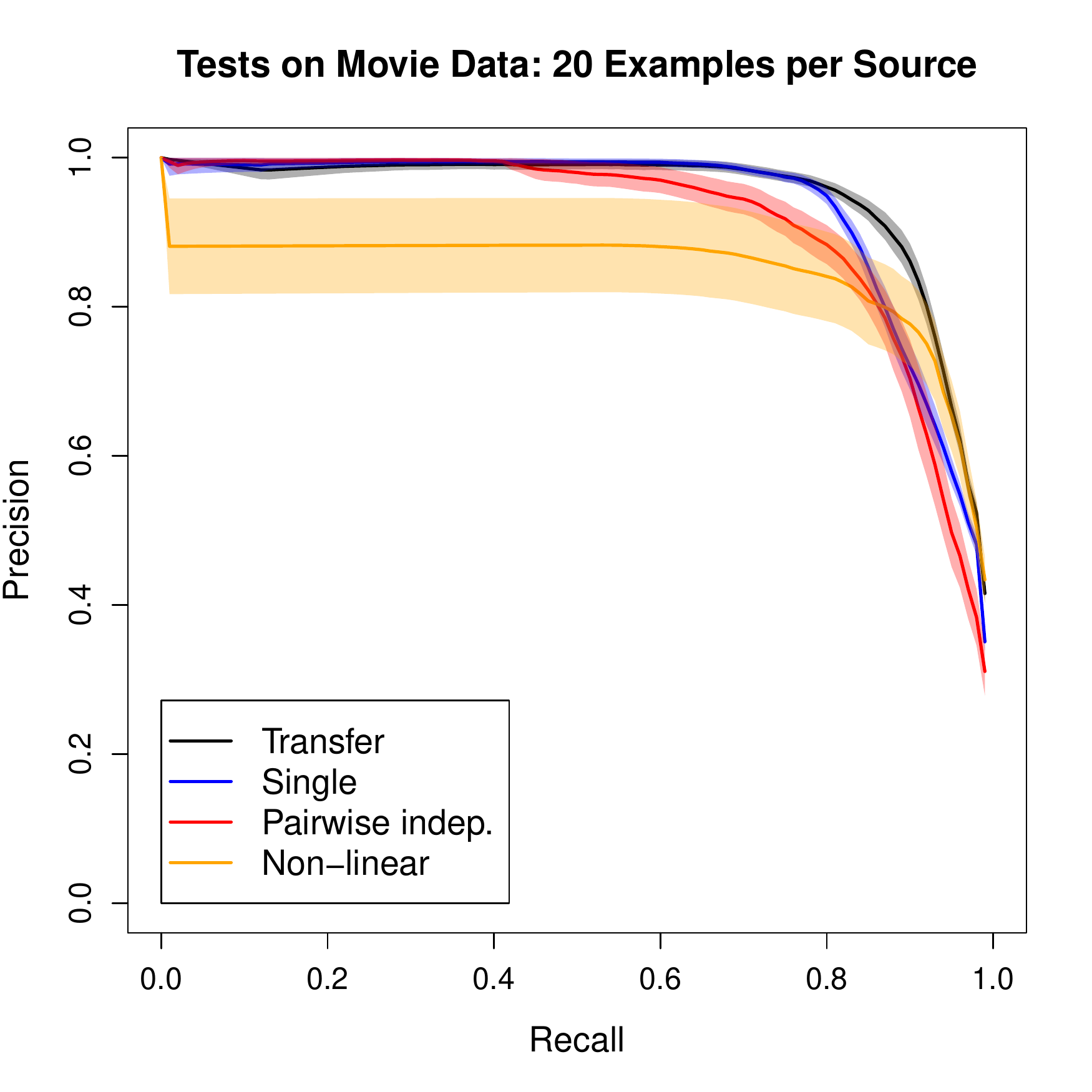} &
\includegraphics[width=.33\linewidth]{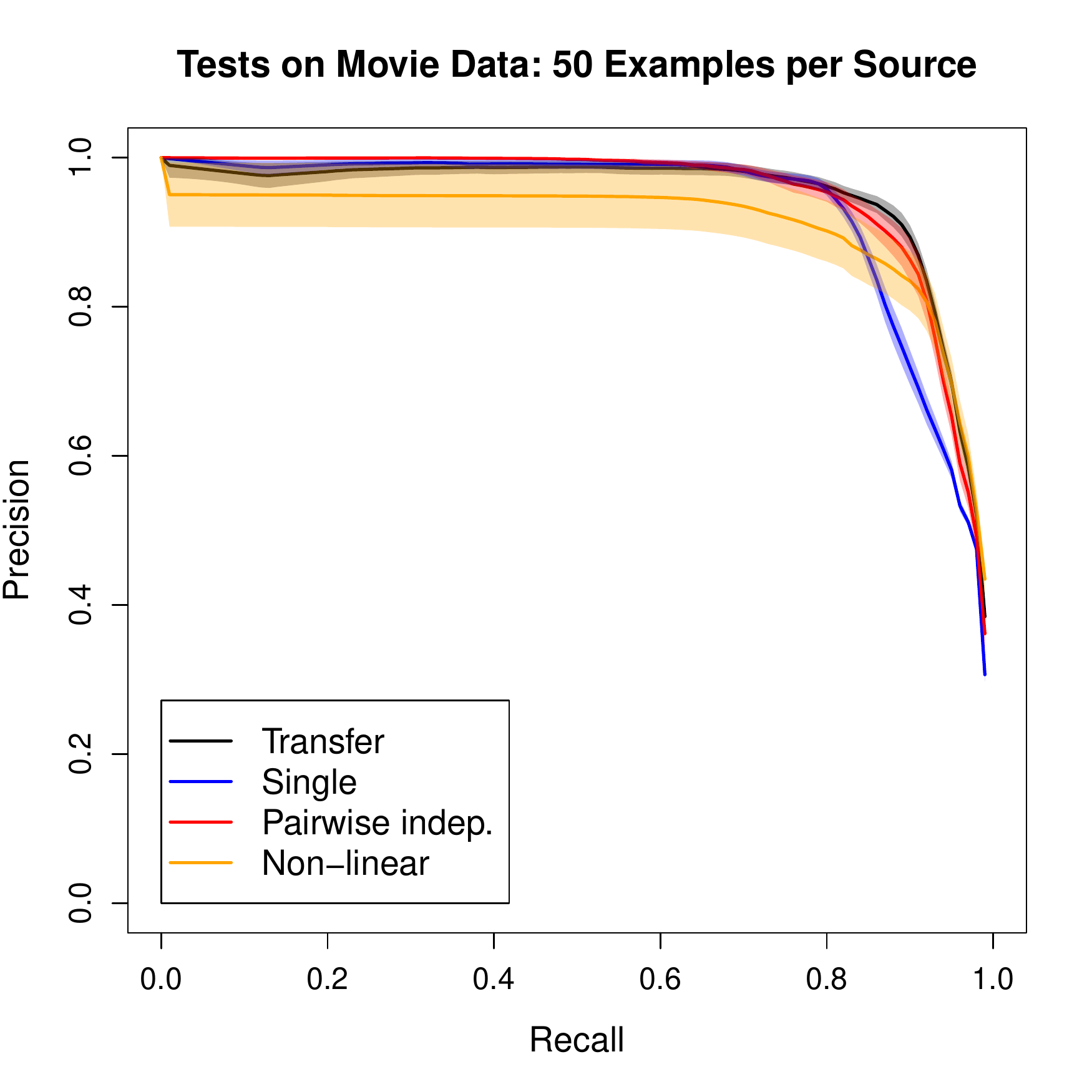}\\
\hline
\includegraphics[width=.33\linewidth]{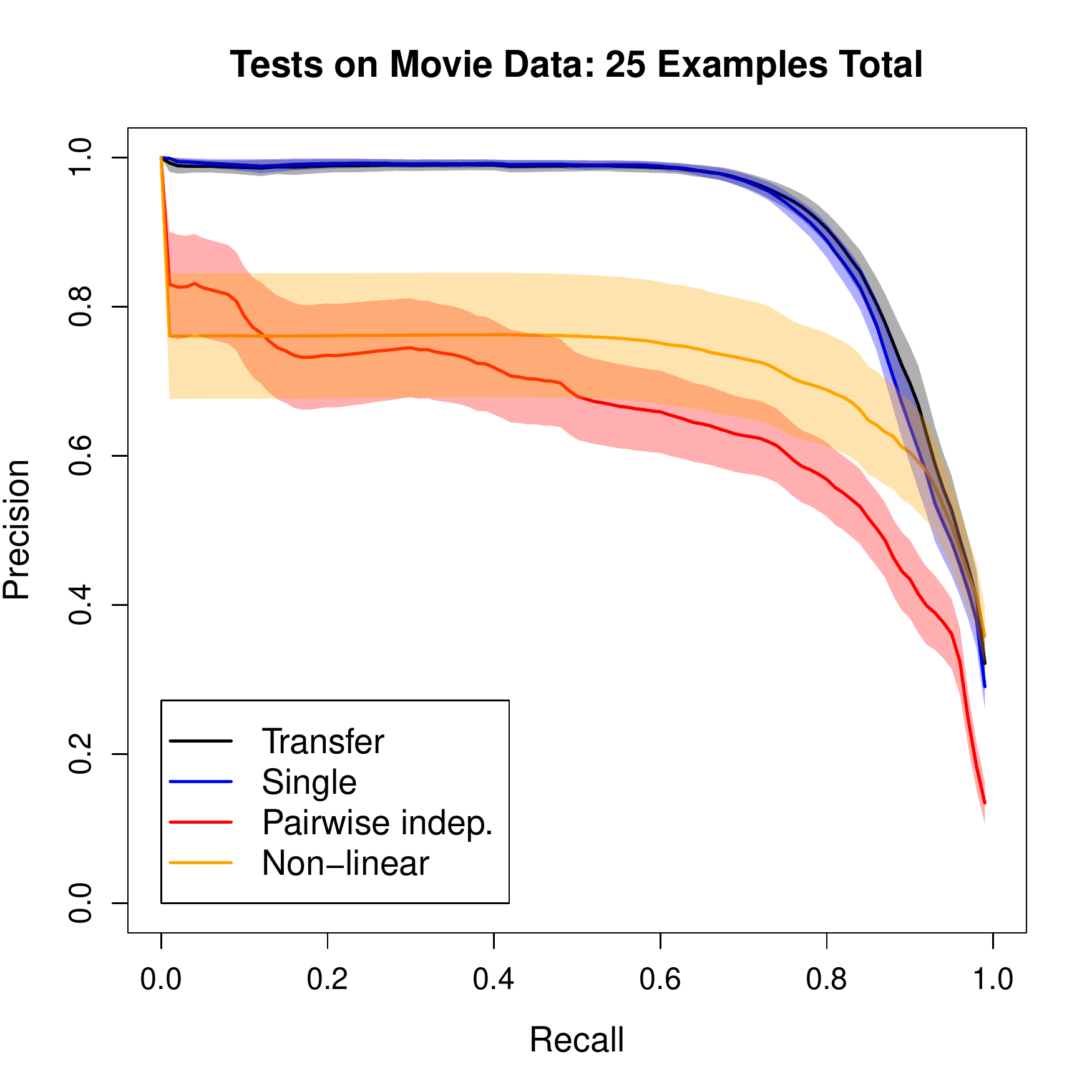} &
\includegraphics[width=.33\linewidth]{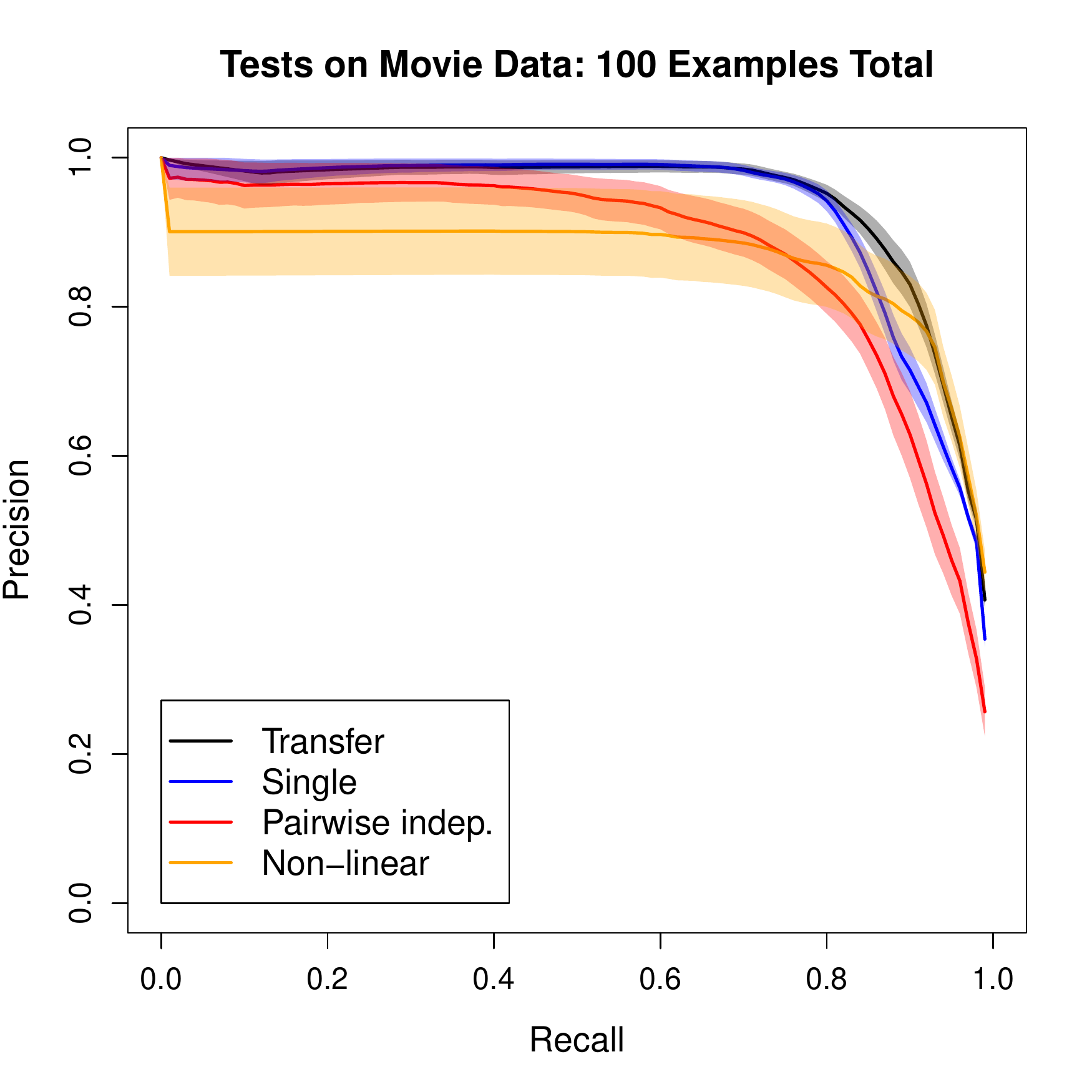} &
\includegraphics[width=.33\linewidth]{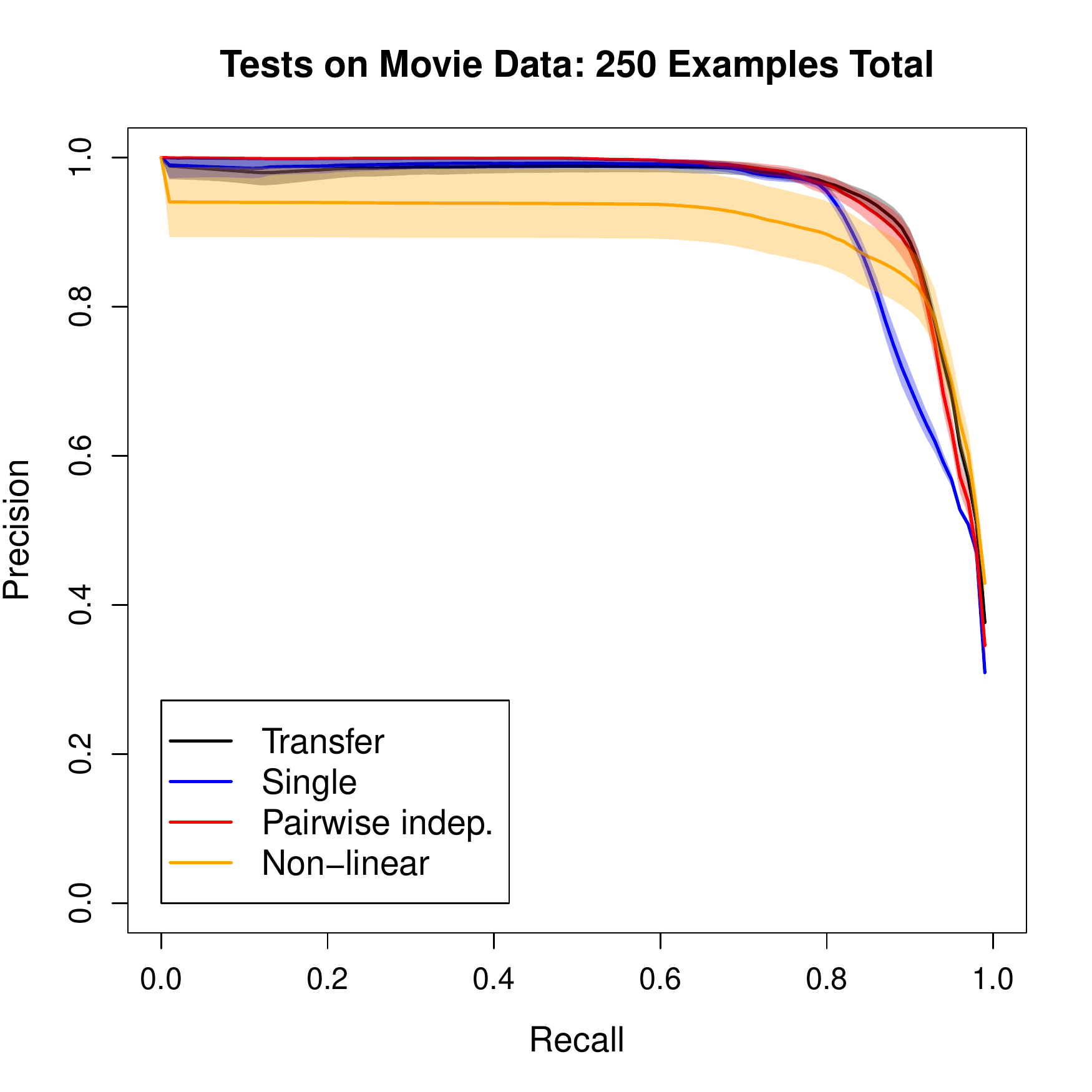}
\end{tabular}
\caption{Precision-recall curves comparing the four learning algorithms on the 6 sources of movie matching data; bands show 95\% pointwise confidence intervals. Figures in the top/middle/bottom row show methods trained on numbers of examples per source pair/source/total; and figures in
columns see increasing training data going from left to right.}
\label{fig:movie-prs}
\end{figure*}

\begin{figure}[t]
\begin{center}
%\begin{minipage}[t]{0.48\textwidth}
\centering
\includegraphics[width=1\linewidth]{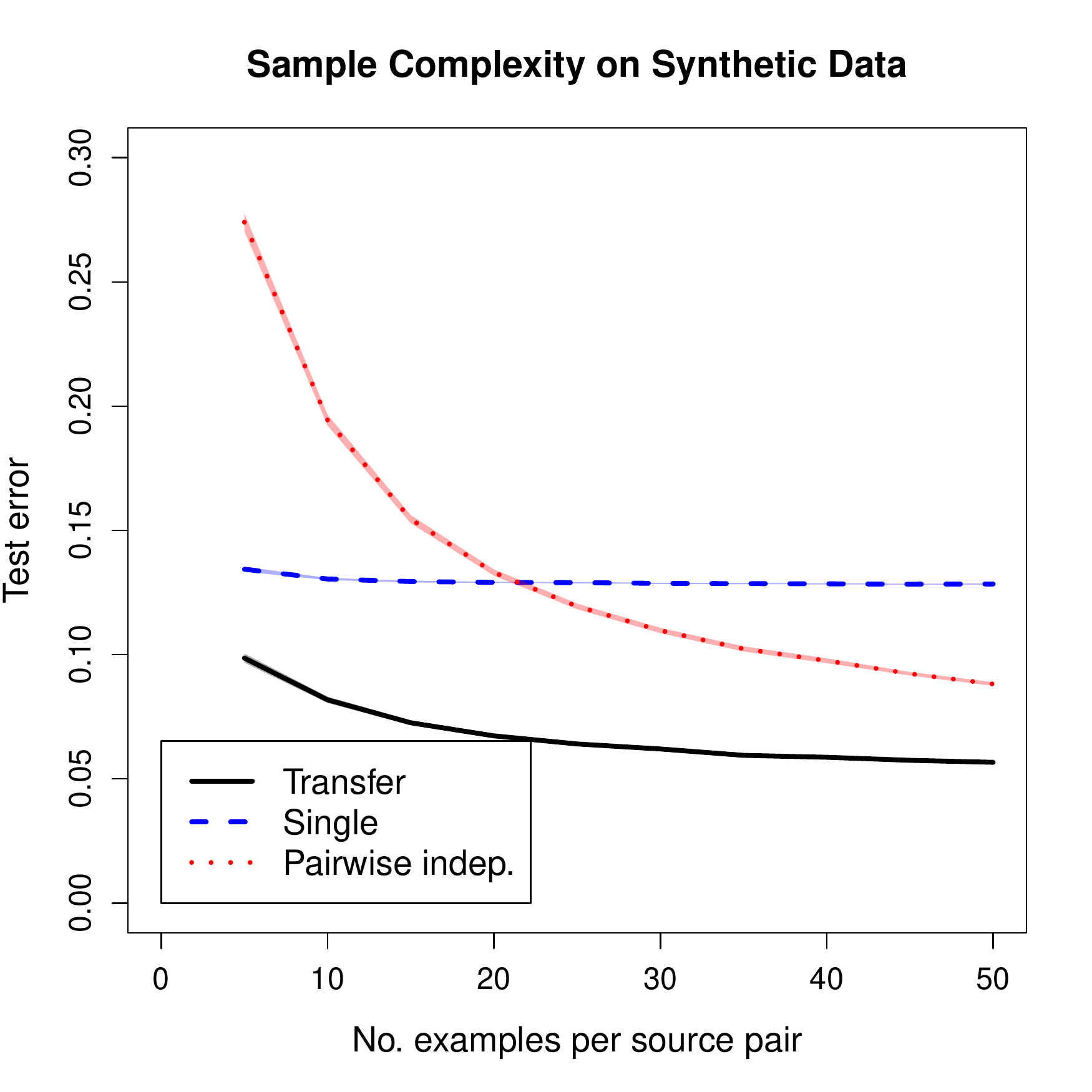}
%\end{minipage}\hfill
%\begin{minipage}[t]{0.48\textwidth}
%\centering
%\includegraphics[width=1\linewidth]{figures/sampleComplexity-trials50-mse.pdf}
%\end{minipage}
\caption{Sample complexity of the three linear learning algorithms on synthetic data.} %Plotting on the left axis (left) test error (right) mean squared error.}
\label{fig:sample-complexity}
\end{center}
\end{figure}

\begin{figure}[t]
\begin{center}
%\begin{minipage}[t]{0.48\textwidth}
%\centering
%\includegraphics[width=1\linewidth]{figures/sourcesComplexity-ex40-trials50-maxSources30-per.pdf}
%\caption{Source complexity of the three learning algorithms on sparse synthetic data.}
%\label{fig:source-complexity-per-sparser}
%\end{minipage}\hfill
\centering
\includegraphics[width=1\linewidth]{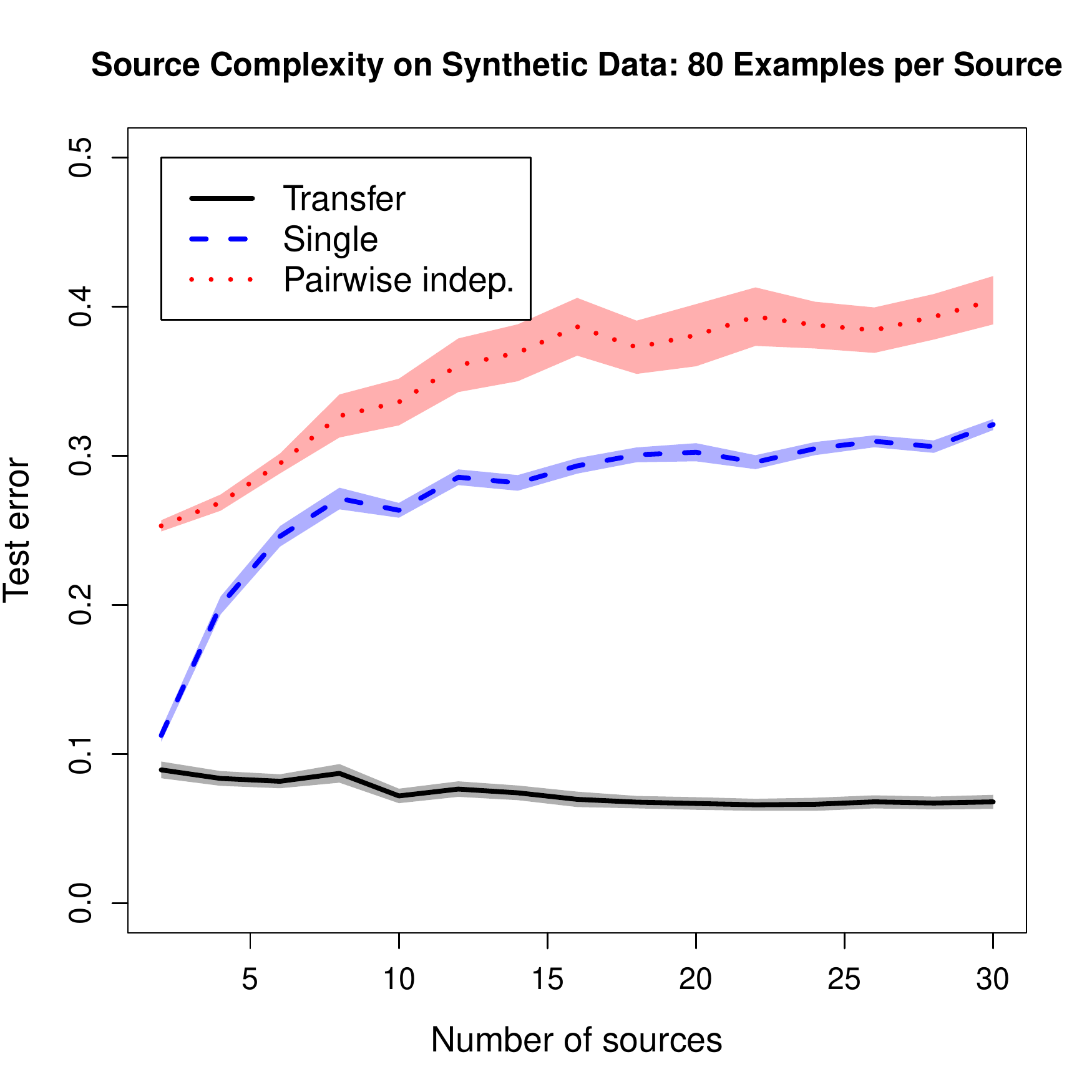}
\caption{Source complexity of the three linear learning algorithms on synthetic data.}
\label{fig:source-complexity-per-denser}
\end{center}
\end{figure}

\section{Results}\label{sec:results}
We now present the results of our experiments, starting on the
movie data. These results are presented
by comparing the PR curves of \transfer and the three baseline methods.
We then focus attention on the synthetic data in order
to gain a deeper understanding of the behavior of the transfer method.
Our results conclusively demonstrate that \transfer requires significantly
less labeling while achieving superior accuracy over state-of-the-art approaches
in multiple-source ER.

%%\lowKeyParagraph{Learned Scoring Functions}
%An interesting aspect of our experiments was that they revealed how
%the sources varied for given attributes. For example, the 
%iTunes database supplies
%wildly varying release dates that were inconsistent with the other
%sources---learned scoring
%functions for all pairs involving iTunes reduced the
%importance on the release year feature.  From that perspective, the
%linear methods supply us with a mechanism to quickly assess various sources
%and their behavior compared to the average behavior of each of the
%individual sources. Such an interpretation can be more difficult
%to obtain from a learned non-linear function.

%The following sections present the results based on using the
%thresholding operator as the classifier. More sophisticated techniques
%can be used to perform the final resolution, however, our
%primary interest in this development is to understand the gains
%achieved in the scoring function. For that reason, we use the
%synthetic data to provide us with a means to more thoroughly explore
%the scores provided by the various learning algorithms.

\subsection{Precision Recall Curves}

With the datasets selected, we applied the four learning algorithm to the
training set, producing pairwise functions $\func_{ij}$.  We then applied the
functions to the unobserved test data and built PR curves by varying threshold
parameter $\thresh$.
Figure~\ref{fig:movie-prs}
presents a three by three grid of PR curves. These figures each
represent the effects of varying the number of training examples on
the precision and recall.
Figure~\ref{fig:prec-vs-labeling} summarizes these nine plots
by fixing recall at 0.85---matching for a movie vertical requires high precision at the expense of
lower recall---and visualizing achieved precision against total number of training
examples.

Consider the summary Figure~\ref{fig:prec-vs-labeling}. As expected \pooled performs
relatively well when very little training data is available, but does not experience much gain
from additional training data---and is inferior to the other methods---owing to it not 
modeling the unique data characteristics of each source. \indep behaves in the opposite
manner to \pooled: it just not able to fit its many-parameter models under little available
data, but progressively improves as more data becomes. \transfer combines the best of
both of the linear baseline models adaptively, and dominates all three 
state-of-the-art baselines at 0.85 recall. While \svm traces the performance of 
\transfer, it is not endowed with the correct pairwise task structure leveraged
by \transfer and so its precision is significantly shifted down.

Similar patterns are born out in the complete PR curves of Figure~\ref{fig:movie-prs} which
are also endowed with 95\% confidence bands.
In its first, second and third rows we vary the number of available examples per source pair,
per source, and in total, respectively. The same trends observed in
Figure~\ref{fig:prec-vs-labeling} are apparent here, however one method does not tend to
dominate another for all recall values.

%In the first row, we vary the number of
%available examples \emph{per source pair}. When the number of examples is
%very small, as shown in the left column, the \indep
%functions perform very poorly and the \svm learner performs
%slightly better while the \pooled and \transfer methods each behave
%roughly equivalently. However, as we increase the number of examples
%per source pair, we see that the PR curves begin to improve, however,
%now, \transfer out-performs all of the other models,
%including the \svm function. As we increase the number of
%examples per source pair, we again see that \transfer 
%performs even better. 
%
%The next row of experiments captures the same
%general behavior, however, we now vary the number of examples \emph{per
%source} and again observe the same general behavior.

%In the final row, we now simply vary the \emph{total} number of examples and
%select training examples uniformly at random from the set of available
%training examples.  In this setting, we again see the same typical
%behavior.
In the final row and column, \indep catches up with
\transfer, because the two sources that were
picked to construct the PR curves are themselves large, so that
a large number of training examples were assigned to that specific
pair. Another interesting observation is that \svm varies 
significantly compared to the other methods (apparent from the width of
the confidence bands). Such behavior is to be
expected as the model complexity for \svm is the greater.
%than for the others. Hence the learning algorithm will
%be able produce more model, however, the learned function
%will be more sensitive to available information and can overfit.
Due to its poor performance on the movie data,
we do not present results for \svm SVM on the synthetic data, where we 
focus on an apples-to-apples comparison of the three linear learners
with varying amounts of transfer.

\begin{figure*}[t]
\begin{center}
\begin{minipage}[t]{0.48\textwidth}
\centering
\includegraphics[width=1\linewidth]{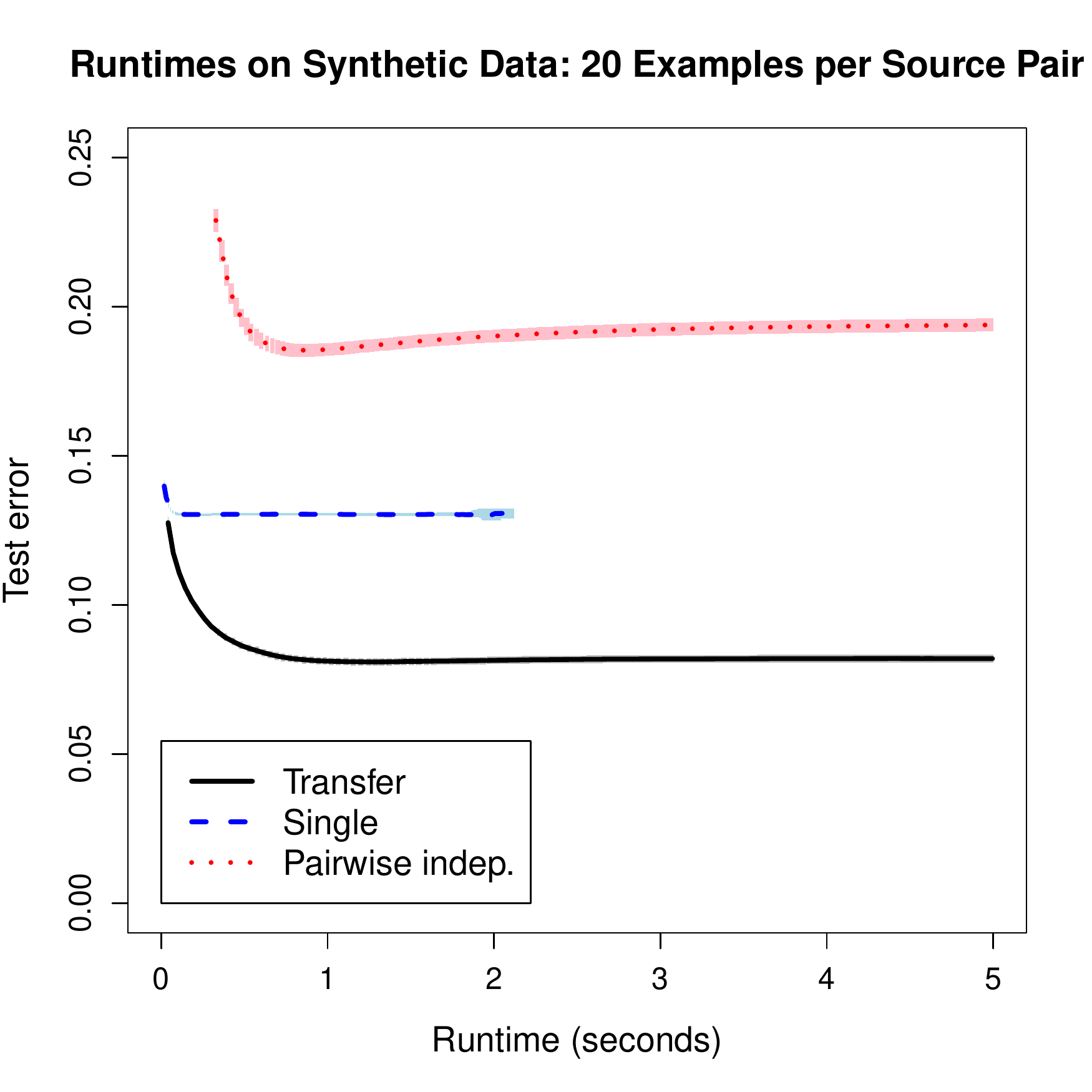}
\end{minipage}\hfill
%\begin{minipage}[t]{0.3\textwidth}
%\centering
%\includegraphics[width=1\linewidth]{figures/timingComplexity-ex30-trials50-per.pdf}
%\end{minipage}\hfill
\begin{minipage}[t]{0.48\textwidth}
\centering
\includegraphics[width=1\linewidth]{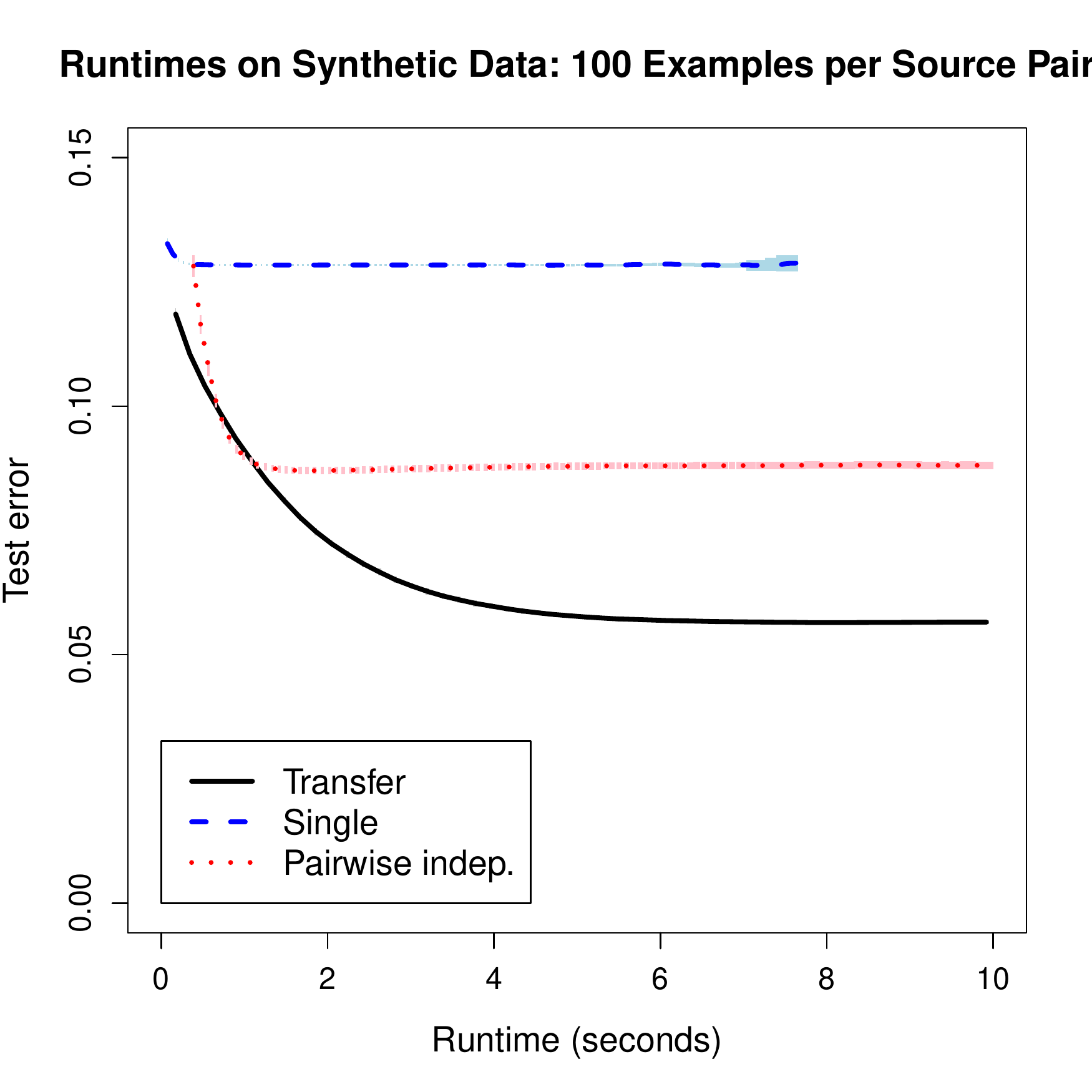}
\end{minipage}
\end{center}
\caption{Runtime analysis of the three learning algorithms on (left) 20 %(middle) 60
(right) 100 examples per synthetic source pair. Across the board transfer yields superior
performance after only a short time. Pairwise independent surpasses pooled once it is
given around 40 examples per source pair.}
\label{fig:timings-per}
\end{figure*}

\subsection{Sample Complexity}

Our next experiment takes a deeper look at understanding the effect of
the number of examples per source pair on the average
error defined in Equation~\eqref{EqProbErr}. These experiments were
performed using synthetic data to give us finer control over
the data generating process and thus concretely explore how increasing
the number of examples will affect $\proberr$.  We observe from
Figure~\ref{fig:sample-complexity}, that as the number of examples
increases \transfer and \indep
both decrease, while \pooled remains lower bounded. This
result is owing to the fact that \pooled cannot take into
account the individual differences between the sources that we are
experimenting with, while the other more flexible methods
can. However, even though \indep does have that freedom, we see that
\transfer still performs better because it is a ``simpler'' model
to learn.

\subsection{Source Complexity}

We now present results of one of our most poignant synthetic experiments.  Figure~\ref{fig:source-complexity-per-denser} shows the
results of increasing the number of synthetic sources from 2 to 30 in
increments of 2 sources. As we increase the number of sources we add a
constant number of training examples---we impose desirable linear not
intractable quadratic labeling cost scaling.
Again we compare the three linear methods,
along with their confidence bands based on 50 trials.
%In the first figure, the synthetic data is produced with sparse feature values, while in the latter we set the synthesis to produce more dense features.
We observe that \transfer achieves far-superior results,
actually improving slightly with the number of sources. On the other
hand, we see that \indep, unsurprisingly, performs
poorly as the number of sources increase as it requires quadratic
scaling of the training data. The algorithm no longer has sufficient available 
examples to train the order of 450 scoring functions.
%for the sparse case we see pooled fairing quite well but for the dense case
We also note that even though \pooled
is very ``simple'', and hence does not require a significant amount of training examples,
it is unable to adapt to the fact that the sources have varying behavior.
Hence, we observe the error increase as the \pooled method is no longer able
to model the behavior of the observed examples.

\subsection{Runtime Analysis}

Figure~\ref{fig:timings-per} shows the results of a timing
analysis of the three linear methods. In the left figure we show the results
with 20 examples per source pair, and in the right we show 100
examples per source pair, both on 10 synthetic sources. In both cases
\transfer quickly achieves superior test error and converges relatively
fast. The baseline linear methods converge faster but to inferior test
error. In the small training set case \pooled outperforms \indep,
and as expected as the number of training examples
increases to 100 examples per source pair, \indep overtakes
\pooled and performs comparably to \transfer.

\section{Related Work}\label{sec:related}

We now discuss prior work in ER as it is related to this paper, from the viewpoints of ER
across sources of varying quality, and recognizing (and mitigating) the cost of generating 
labeled data. This paper is one of the first to consider ER on multiple sources of varying quality,
is the first to highlight cost of labeling as a barrier to scaling \emph{accurate} ER to multiple
sources, and is the first to apply the transfer learning paradigm to ER (and along the way
we develop a transfer learning algorithm that is novel to the machine learning community).

\lowKeyParagraph{Varying Source Quality}
Numerous works have studied ER, many of which involving matching
across multiple sources, but very few have explored the serious challenge of resolving
sources with varying data quality. Traditionally researchers use a single non-learning-based
matcher or model a single learner on pooled training data~\cite{SourceAware07}.
Shen \etal posited that real-world data sources have varying levels of \term{semantic ambiguity},
a special case of source quality in which an individual attribute should be given more or less weight.
They propose the SOCCER framework for compiling matching execution plans in which one of two
matchers can be employed on different sources---a relaxed (conservative) matcher requiring less
(respectively more) evidence to declare a match. These matchers could be related via the relaxation
of a threshold, for example. Similar to~\cite{SourceAware07}, the adverse effect of heterogeneous
source quality on matching accuracy is a key motivation of this paper. By contrast, however, our
paper sets out to \emph{learn continuous characterizations} of quality via real-valued weights,
over \emph{all} attributes together. Moreover we leverage a significantly more fine-grained transfer
structure between the tasks of matching different source pairs, compared to the authors' process of
simply pooling like tasks---which is more akin to the straw man \pooled learner used for comparison
here, which does not correctly balance transfer with the needs of matching tasks.

%In work parallel to this research, \singleDoubleBlind{we have previously}{Bo \etal have} explored approaches to merging attributes of already-matched entities
%where entity sources have varying quality~\cite{LTM12}. Like the present work, their motivation to model
%source quality is borne out of the recognition that heterogeneous sources can negatively impact resolution quality.
%However there the similarity ends: while this paper presents a supervised frequentist approach to transfer
%relative patterns on feature-level data between different similarity learning tasks, \cite{LTM12} presents a
%Bayesian probabilistic graphical model---Latent Truth Model---which infers source quality parameters for each individual source,
%for a single attribute such as book author or movie director. The method is unsupervised in that it does not require human
%labeling, however in a sense it is supervised by the output of matching. Finally LTM does not employ any kind of transfer learning
%between learning tasks.

K\"opcke \& Rahm developed the STEM framework for investigating the effect of training selection on learning to match~\cite{QDBMUD08-STEM}.
While they do not compare training on heterogeneous sources independently in a pairwise fashion versus together with 
a single matcher, they implicitly acknowledge the need to produce different matchers tailored to source-pairs' characteristics
as they split their most challenging experimental matching task of resolving publications between three sources---Google Scholar (larger but of lower quality),
DBLP and the ACM Digital Library (both higher quality but smaller)---into independent learning problems between DBLP and the other two sources.

\lowKeyParagraph{The Cost of Labeling} The key challenge solved by our approach is to significantly reduce the human effort 
required to label training data for learning to combine matchers over multiple domains. In their thorough comparative evaluations of 21
ER systems with their FEVER framework~\cite{Kopcke2010a}, K\"opcke \etal explicitly identified human effort as a key metric for the
effectiveness of learning-based ER systems. In their line of work, this desire can be traced to their earlier paper on STEM~\cite{QDBMUD08-STEM}. In both works the authors pay particular attention to the effect of training-set size on the quality of matching, and favor methods
requiring less labeled data such as the SVM. Unlike this paper, however, they do not consider matching across multiple sources and the
additional requirement this can place on human labeling.

While low sample complexity has been identified as a desirable property of entity matchers, the notion of labeling cost has not been previously recognized as a barrier to scaling ER. We introduce the notion of \term{source complexity} which characterizes the change in error as new sources are added,
provided the number of training examples are increased only linearly with the sources.

\section{Conclusions}\label{sec:conc}

Many problems in databases, statistics and machine learning require learning of a pairwise similarity function 
from human-labeled examples. However as the number of data sources increases, the sample complexity---the cost of human labeling---increases quadratically. To overcome this prohibitive scaling, we propose a new transfer learning algorithm \transfer for learning multiple similarity score functions jointly. We take ER as a motivating example, and present extensive experimental comparisons of \transfer against existing state-of-the-art methods. Our experiments---on real-world, large scale movie matching data, and extensive synthetic data---show that \transfer indeed produces more accurate results for ER than existing methods, with less data, and indeed in faster time. Interesting future work might consider combining
active learning with \transfer, and extending \transfer to non-linear classification.

\singleDoubleBlind{\section{Acknowledgments}

We thank Ashok Chandra, Ariel Fuxman, Duo Zhang and Bo Zhao for their valuable comments and assistance.
}{}

\bibliographystyle{abbrv}
\bibliography{sources}  % sigproc.bib is the name of the Bibliography in this case
% ACM needs 'a single self-contained file'!
%\balancecolumns
%\appendix
%Appendix A
%\section{Headings in Appendices}

%\balancecolumns % GM June 2007
% That's all folks!
\end{document}